\documentclass[a4paper,11pt,nohyper]{JHEP3}
\def\d{\delta}

\def\t{\tau}

\def\S{\Sigma}

\def\del{\partial}              
   \let\d=\delta

 \let\t=\tau

\def\nn{\nonumber} \def\bd{\begin{document}} \def\ed{\end{document}}
\def\ds{\documentstyle} \let\fr=\frac \let\bl=\bigl \let\br=\bigr
\let\Br=\Bigr \let\Bl=\Bigl
\let\bm=\bibitem
\let\na=\nabla
\let\pa=\partial \let\ov=\overline
\newcommand{\be}{\begin{equation}}
\newcommand{\ee}{\end{equation}}
\def\ba{\begin{array}}
\def\ea{\end{array}}
\def\ft#1#2{{\textstyle{{\scriptstyle #1}\over {\scriptstyle #2}}}}
\def\fft#1#2{{#1 \over #2}}
\def\del{\partial}
\def\sst#1{{\scriptscriptstyle #1}}
 \def\oneone{\rlap 1\mkern4mu{\rm l}}
\def\ie{{\it i.e.\ }}
\def\via{{\it via}}
\def\semi{{\ltimes}}
\def\str{{\rm str}}
\def\Dm{{{D_{\sst{max}}}}}
\def\vac{ \left | 0 \right \rangle }
\def\kvac{ \left | k \right \rangle }

\def\sp{\; \; \;}

\def\bol{ \left | B (p^+) \right \rangle}
\def\bo1{ \left | B^0 (p^+) \right \rangle}

\def\bolt{ \left | B (p^+) \right \rangle_{\t}}

\def\boxl{ \left | B (x^-) \right \rangle}
\newcommand{\bea}{\begin{eqnarray}}
\newcommand{\eea}{\end{eqnarray}}

\def\<{ \langle }
\def\>{ \rangle }

\def\S{\Sigma}

\usepackage{amsmath,latexsym,amssymb,slashed}
\usepackage{graphicx}
\usepackage{psfrag}
\usepackage[latin1]{inputenc}
\usepackage{subfigure}

\renewcommand{\floatpagefraction}{0.6}
\renewcommand{\textfraction}{0.2}

\newcommand\ca{\mathcal{A}}
\newcommand\vp{\varphi}
\newcommand\beal{\begin{align}}
\newcommand\bbone{\ensuremath{\mathbbm{1}}}

\newcommand{\eq}[1]{\begin{equation}#1\end{equation}}
\newcommand{\spl}[1]{\begin{split}#1\end{split}}
\newcommand{\al}[1]{\begin{align}#1\end{align}}
\newcommand{\subeq}[1]{\begin{subequations}#1\end{subequations}}

\newcommand{\arXividhepth}[1]{\href{http://arxiv.org/abs/#1}arXiv:{\tt #1} [hep-th]}
\newcommand{\arXividother}[2]{\href{http://arxiv.org/abs/#1}arXiv:{\tt #1} [#2]}

\newcommand{\bg}[1]{\hat{#1}}
\newcommand{\wj}{\widetilde{J}}
\newcommand{\reo}{\mathrm{Re}~\!\omega}
\newcommand{\imo}{\mathrm{Im}~\!\omega}
\newcommand{\ads}{AdS_4}
\newcommand{\mcal}{\mathcal{M}}
\newcommand{\ccal}{\mathcal{C}}
\newcommand{\ncal}{\mathcal{N}}
\newcommand{\boxedeq}[1]{
\begin{equation}
\fbox{
\rule[0.7cm]{0pt}{0pt}
$#1$
\rule[-0.45cm]{0pt}{0pt}
}
\end{equation}
}

\def\d{\text{d}}

\def\slashchar#1{\setbox0=\hbox{$#1$}           
\dimen0=\wd0                                 
\setbox1=\hbox{/} \dimen1=\wd1               
\ifdim\dimen0>\dimen1                        
\rlap{\hbox to \dimen0{\hfil/\hfil}}      
#1                                        
\else                                        
\rlap{\hbox to \dimen1{\hfil$#1$\hfil}}   
/                                         
\fi}

\def\Re           {{\rm Re\hskip0.1em}}
\def\Im           {{\rm Im\hskip0.1em}}

\newcommand{\E}{\text{\tiny E}}

\newcommand{\tV}{{\widetilde{V}}}
\newcommand{\tH}{{\tilde{h}}}
\newcommand{\tm}{{{m}}}
\newcommand{\tmu}{{\tilde{\mu}}}
\newcommand{\trho}{{\tilde{\rho}}}
\newcommand{\tv}{{\tilde{v}}}
\newcommand{\calo}{\mbox{${\cal O}$}}
\newcommand{\cala}{\mbox{${\cal A}$}}
\newcommand{\dd}{\mathrm{d}}
\newcommand{\ra}{\rightarrow}
\newcommand{\calv}{\mbox{${\cal V}$}}
\newcommand{\calh}{\mbox{${\cal H}$}}
\newcommand{\calm}{\mbox{${\cal M}$}}
\newcommand{\abs}[1]{\left| #1 \right|}
\newcommand{\zetaa}{{\psi}}
\newcommand{\tr}{{\rm tr}\,}

\newcommand{\ky}[1]{{\color{blue}{#1}}}

\title{Renormalized entanglement entropy}

\author{Marika Taylor and William Woodhead \\

Mathematical Sciences and STAG Research Centre, University of Southampton, \\
Highfield, Southampton, SO17 1BJ, UK.

\bigskip
 E-mail:
 \email{m.m.taylor@soton.ac.uk; w.r.woodhead@soton.ac.uk}}

\abstract{We develop a renormalization method for holographic entanglement entropy based on area renormalization of entangling surfaces. 
The renormalized entanglement entropy is derived for entangling surfaces in asymptotically locally anti-de Sitter spacetimes in general dimensions and
for entangling surfaces in four dimensional holographic renormalization group flows. The renormalized entanglement entropy for disk regions in $AdS_4$ 
spacetimes agrees precisely with the holographically renormalized action for $AdS_4$ with spherical slicing and hence with the F quantity, 
in accordance with the Casini-Huerta-Myers map. We present a generic class of holographic RG flows associated with deformations by operators of dimension
$3/2 < \Delta < 5/2$  for which the F quantity increases along the RG flow, hence violating the strong version of the F theorem. We conclude by explaining
how the renormalized entanglement entropy can be derived directly from the renormalized partition function using the replica trick i.e. our renormalization method
for the entanglement entropy is inherited directly from that of the partition function. We show explicitly how the entanglement entropy 
counterterms can be derived from the standard holographic renormalization counterterms for asymptotically locally anti-de Sitter spacetimes.} 

\begin{document}

\newcommand{\td}{\tilde}
 \newcommand{\bc}{\begin{center}}
 \newcommand{\ec}{\end{center}}
 \newcommand{\bfr}{\begin{flushright}}
 \newcommand{\efr}{\end{flushright}}
 \newcommand{\bfl}{\begin{flushleft}}
 \newcommand{\efl}{\end{flushleft}}
 \newcommand{\bt}{\begin{tabular}}
 \newcommand{\et}{\end{tabular}}

\section{Introduction}

In recent years there has been considerable interest in entanglement entropy and its holographic implementation, following the proposal of
\cite{Ryu:2006bv} that entanglement entropy can be computed from the area of a bulk minimal surface
homologous to a boundary entangling region. This proposal was proved for spherical entangling regions in conformal field theories in \cite{Casini:2011kv} and
arguments supporting the Ryu-Takayanagi prescription based on generalized entropy were given in \cite{Lewkowycz:2013nqa}.  Entanglement entropy has by now been computed in a wide range of holographic systems, see the review \cite{Takayanagi:2012kg}. General properties of holographic entanglement entropy are reviewed in 
\cite{Headrick:2013zda}. 

Entanglement entropy is a UV divergent quantity, with the leading UV divergences scaling with the area of the boundary of the entangling region. For a quantum field theory in D spatial dimensions, the boundary of the entangling region is $(D-1)$-dimensional and thus $
S \approx \Lambda^{D-1} {\cal A}_{D-1} $
where $\Lambda$ is the UV cutoff and ${\cal A}_{D-1}$ is the area of the boundary of the entangling region. 

If one is interested in the entanglement entropy of a discrete system, in which there is a natural UV cutoff set by, for example, the lattice scale, then it may be natural to work with this ``bare" entanglement entropy. If however one is interested in entanglement entropy in a quantum field theory context, then it natural to explore whether and how entanglement entropy can be renormalized. 

Finite terms in the entanglement entropy are used in a number of contexts. Firstly, they arise as order parameters for phase transitions, see the pioneering  works \cite{Klebanov:2007ws,Huijse:2011ef}. Finite terms in the entanglement entropy for disk regions in three dimensional conformal field theories are also related by conformal transformations \cite{Casini:2011kv} to the free energy on a three sphere, which is the quantity appearing in the proposed F theorem \cite{Jafferis:2011zi}. 

As we will review in section \ref{two}, in previous works the finite terms in the entanglement entropy have been isolated using differentiation of the entanglement entropy with respect to geometric parameters characterizing the entangling region. Such procedures can be implemented in a simple way, both holographically and in field theory calculations, but they have several disadvantages. The differentiation prescriptions depend on the specific geometry of the entangling region, and thus it is hard to implement such renormalization in situations where the shape of the entangling region is itself being varied. Renormalization by differentiation is furthermore not directly related to the renormalization procedures used for other quantum field theory quantities. Thus, in particular, it is hard to understand issues such as the scheme dependence of the finite answer. 

In this paper we will develop a systematic renormalization procedure for entanglement entropy. We begin by setting up holographic renormalization for the Ryu-Takayanagi entanglement entropy functional. Since the entanglement entropy is described by the area of a minimal surface homologous to the boundary entangling region, the UV divergences of the entanglement entropy are in direct correspondence with the area divergences of this minimal surface. Following the holographic renormalization methods of \cite{Henningson:1998gx,Henningson:1999xi,DeHaro2001} one can identify covariant counterterms on the conformal boundary of the minimal surface which renormalize the area of the minimal surface. 

In section \ref{three} we derive the renormalized Ryu-Takayanagi functional for static entangling surfaces in AdS spacetimes. Assuming flat spatial slices of the background for the dual quantum field theory (i.e. a Poincar\'{e} representation of $AdS_{D+2}$) the renormalized functional takes the form
\bea
S_{\rm ren} &=& \frac{1}{4 G_{D+2}} \int_{\Sigma} \dd^D \sigma^{\alpha} \, \sqrt{ \gamma}  \\ 
&& \qquad - \frac{1}{4G_{D+2}}  \int_{\del \Sigma} \dd^{D-1} x \sqrt{\tilde{\gamma}}
\left (  \frac{1}{D-1} + \frac{1}{2(D-1)^2(D-3)} {\cal K}^2 \cdots \right ), \nn
\eea
Here $\Sigma$ is the entangling surface,with induced metric $\gamma$, and $\partial \Sigma$ is its boundary, with induced metric $\tilde{\gamma}$. The extrinsic curvature ${\cal K}$ refers to the extrinsic curvature of $\partial \Sigma$ embedded into a spatial slice of the boundary of the bulk manifold. 
The first counterterm becomes logarithmic for $D=1$. Only the first
counterterm given above is needed for gravity in four bulk dimensions ($D=2$). The second counterterm becomes logarithmic at $D=3$ and is needed in the form given above for $D > 3$. Additional counterterms involving higher order curvature invariants are needed for $D \ge 5$. The counterterms for entangling surfaces in general asymptotically locally $AdS$ spacetimes can be found in section \ref{five}. 

We then show that the renormalized entanglement entropy for a disk region in a three dimensional conformal field theory dual to $AdS_4$ is in precise agreement with the holographically renormalized Euclidean action for $AdS_4$ with spherical slicing, i.e. the CHM map \cite{Casini:2011kv} holds at the level of renormalized quantities. 

\bigskip

In section \ref{four} we consider holographic RG flows in four bulk dimensions which respect Poincar\'{e} invariance of the dual theory. For flows driven by a single scalar we compute the renormalized Ryu-Tayakanagi functional, expressing the counterterms in terms of the superpotential associated with the flow. 

We then use the renormalized entanglement entropy to explore the change in the F quantity along RG flows. In particular, we consider a disk entangling region and calculate the change the renormalized entanglement entropy (and hence F quantity) perturbatively in the source of the relevant deformation, $\phi_{(0)}$. For operators of dimension 
$3/2 < \Delta_{+} < 3$ we find that 
\be
\delta S_{\rm ren} =  \frac {\pi }{16 (2 \Delta_+ - 5) G_4} \phi_{(0)}^2 R^{2 (3- \Delta_+)} + {\cal O} \left ( \phi_{(0)}^3 \right ),
\ee
where $R$ is the radius of the disk entangling region while $\delta S_{\rm ren} = 0$ for exactly marginal operators. This quantity is clearly negative for $\Delta_{+} < 5/2$ which, since $\delta S_{\rm ren } = - \delta F$, corresponds to an increase in the F quantity. We should note however that the corresponding deformations on the three sphere are inhomogeneous and do not therefore correspond to RG flows which respect the $SO(4)$ invariance.  
Direct calculation of the F quantity for $SO(4)$ invariant RG flows on $S^3$ driven by such operators also gives an increase in the F quantity to quadratic order in the source, see the companion paper \cite{Taylor1}. It would be interesting to understand whether such flows are unphysical or if the strong version of the proposed F theorem is indeed violated.

\bigskip

In section \ref{five} we show that the holographically renormalized entanglement entropy can be obtained from the holographically renormalized action. Using the replica trick, 
the entropy associated with a density matrix $\rho$ is expressed as
\be
S = -n \partial_n \left [ \log Z(n) - n \log Z(1) \right ]_{n=1} \label{rep1a}
\ee
where $Z(n) = {\rm Tr} (\rho^n)$ and $Z(1)= {\rm Tr} (\rho)$ is the usual partition function. If we are interested in the entropy of a thermal state, then $Z(n)$ is constructed by extending the period of the thermal circle by a factor of $n$. In the case of entanglement entropy, $Z(n)$ is constructed by extending the period of the circle around the boundary of the entangling region by a factor of $n$, where implicitly $n$ is an integer. Assuming that the resulting expression is analytic in $n$, one can obtain the entropy by analytically continuing to $n=1$. 

Holographically $Z(n)$ can be computed in terms of the onshell Euclidean action \cite{Lewkowycz:2013nqa} as
\be
S = n \partial_n \left [ I(n) - n I(1) \right ]_{n=1}. \label{rep2a}
\ee
Here $I(1)$ represents the onshell Euclidean action for the bulk geometry while $I(n)$ represents the onshell Euclidean action for the replica bulk geometry.
For a thermal state, the bulk geometry associated with $Z(1)$  is a black hole and the replica is constructed by extending the period of the thermal circle by a factor of $n$. 
For the entanglement entropy, the bulk geometry associated with $Z(1)$ corresponds to the usual bulk dual of the given state in the field theory and
the replica is constructed by extending the period of the circle around the entangling region boundary by a factor of $n$. Following the same logic as in Lewkowycz-Maldacena
 \cite{Lewkowycz:2013nqa}, the expression \eqref{rep2a} localises on the minimal surface corresponding to the extension of the boundary of the entangling region into the bulk. However,  the entangling surface itself has area divergences, unlike the black hole setup analysed in detail in  \cite{Lewkowycz:2013nqa}. 

In section \ref{five} we show that the renormalized entanglement entropy can be expressed in terms of the renormalized onshell action i.e. 
\be
S_{\rm ren} = n \partial_n \left [ I_{\rm ren}(n) - n I_{\rm ren}(1) \right ]_{n=1} \label{rep3a}.
\ee
In particular, using the standard counterterms for asymptotically locally $AdS$ spacetimes \cite{DeHaro2001}, together with results on the curvature invariants of the replica space
\cite{Solodukhin:2008dh,Fursaev:2013fta}, one obtains exactly the same $S_{\rm ren}$ as computed directly via area renormalization. Thus, the renormalization scheme for the entanglement entropy is inherited directly from the renormalization scheme used for the partition function. 

This result provides evidence for the applicability of the replica trick in the holographic context. Note that the derivation of the entanglement entropy functional from the Euclidean action functional requires only the local geometry of the replica; any potential anomalies in the replica symmetry do not affect the derivation. The holographic renormalization counterterms for higher derivative gravity theories such as Gauss-Bonnet also imply counterterms for the entanglement entropy, as we discuss at the end of section \ref{five}. 

\bigskip

The plan of this paper is as follows. In section \ref{two} we review the renormalization of entanglement entropy by differentiation. In section \ref{three} we setup area renormalization for entangling surfaces in $AdS$ spacetimes, and show that the renormalized entanglement entropy for disk regions in $AdS_4$ indeed agrees with the F quantity. In section \ref{four} we consider entanglement entropy for RG flows while in section \ref{five} we show how the renormalized entanglement entropy can be obtained from the renormalized action via the replica trick. We conclude in section \ref{six}.

\section{Renormalization by differentiation} \label{two}

In previous works, the finite terms in the entanglement entropy have been isolated by differentiation of the entanglement entropy. In the case of a strip of width $R$, UV divergent contributions to the entanglement entropy in a local quantum field theory are necessarily independent of $R$ and therefore 
\begin{equation}
S_R = R \frac{\partial S}{\partial R}
\end{equation}
is finite. This expression has been used in a number of earlier works, including \cite{Casini:2005zv,Casini:2006hu,Calabrese:2009qy}. 

For a spherical entangling region, the radius of the sphere controls the local curvature of the boundary of the entangling region and therefore it is no longer true that UV divergences are independent of the scale of the entangling region. In \cite{Liu:2012eea} it was noted that the following quantity 
\begin{equation}
F(R) = - S(R) + R \frac{\partial S}{\partial R} \label{sr}
\end{equation}
is manifestly finite in any 3d field theory which has a UV fixed point. (Analogous expressions for general dimensions were given in  \cite{Liu:2012eea}.) In particular, for a three-dimensional CFT the regulated entanglement entropy for a disc entangling region is 
\be
S_{\rm reg} = \frac{a_{-1} R}{\delta} + a_{0} \label{div-r}
\ee
where $\delta \ll 1$ is the UV cutoff and $(a_0,a_{-1})$ are constants. Then by construction
\be
F(R) = - a_0.
\ee
For theories with a holographic dual one can show (see section \ref{three}) that 
\be
S_{\rm reg} = \frac{\pi}{2 G_4} \left ( \frac{R}{\delta} - 1 \right )
\ee
and therefore 
\be
F(R) = \frac{\pi}{2 G_4}.
\ee
The normalization of \eqref{sr} is chosen so that the latter indeed agrees with the F quantity. 

The renormalized entanglement entropy defined by (\ref{sr}) has both positive and negative features. On the positive side, there is evidence that $F(R)$ behaves monotonically as a function of $R$ in free field theory and holographic examples \cite{Klebanov:2012yf,Liu:2013una}. Also by construction 
\be
\frac{\partial F}{\partial R} = R \frac{\partial^2 S}{\partial R^2}
\ee
and strong subadditivity of the entanglement entropy implies that in any Poincar\'{e} invariant field theory $\partial^2 S/\partial R^2 \le 0$ \cite{Casini:2012ei}, so $F(R)$ is a non-increasing function of the radius $R$. 

 Let us suppose we deform a conformal field theory by an operator ${\cal O}_{\Delta}$ of dimension $\Delta <  3$:
\be
I_{CFT} \rightarrow I_{CFT} + \int d^3 x \sqrt{h}  \lambda {\cal O}_{\Delta}.
\ee
The dimension of $\lambda$ is then $(3 - \Delta)$; the coupling provides another dimensionful scale and it is no longer the case that \eqref{div-r} are the only divergences. There are in general additional divergences which are analytic in the deformation parameter $\lambda$ and hence for a disk region the change in the entanglement entropy under the relevant deformation is 
\be
\delta S_{\rm reg} = a_{5 - 2 \Delta} \frac{\lambda^2 R}{\delta^{\Delta - 5/2}} + a_{8 - 3 \Delta} \frac{\lambda^3 R}{\delta^{\Delta - 8/3}} + \cdots 
\ee
where the coefficients $a_{m}$ are dimensionless. Hence for $\Delta > 5/2$ the relevant deformation generates additional UV divergences in the entanglement entropy; additional divergences arise for $\Delta > 3 - 1/n$. The form of this expression follows from conformal perturbation theory; in particular the term linear in $\lambda$ vanishes, while all divergences scale extensively with the length of the boundary of the entangling region.  By construction $F(R)$ is finite for all such deformations although it is not a priori clear that $F(R)$ agrees with the F quantity. 

On the negative side, there is evidence that $F(R)$ is not stationary at a UV fixed point \cite{Klebanov:2012va}. Consider perturbations of a  two-dimensional CFT by a slightly relevant operator of dimension $2 - \delta_{\Delta}$. Then Zamoldchikov's c-function behaves as
\be
c(g) = c_{UV} - g^2 \delta_{\Delta} + {\cal O}(g^3)
\ee
where $g$ is the renormalised coupling. For a theory with several coupling constants
\be
\frac{\partial c}{\partial g^i} = G_{ij } \beta^j
\ee
where $G_{ij}$ is the Zamalodchikov metric and $\beta^j = \mu \frac{\partial g^j}{\partial \mu }$ are the beta functions. Then non-singularity of the Zamalodchikov metric guarantees the stationarity of the $c$ function in two dimensions. In  \cite{Klebanov:2012va} it was shown that the proposed $F(R)$ is not stationary in this sense at the UV fixed point in free massive scalar field theory examples. 

Another drawback of the definition of the renormalized entanglement entropy \eqref{sr} is that the definition is only applicable to disk entangling regions, or to regions which are characterized by one overall scale. This drawback is not an issue for applications to the F theorem, for which only disk regions are needed, but prevents using \eqref{sr} to explore the general shape dependence of entanglement entropy. 

The renormalization that we propose in this paper by contrast is inherited directly from the renormalization of the partition function, making scheme dependence and the relation to the F quantity manifest, and is applicable to any shape entangling region. Moreover, our renormalization is applicable in theories which are not conformal in the UV.  

\section{Renormalized entanglement entropy in anti-de Sitter} \label{three}

In this section we will define the renormalized area of (static) entangling surfaces in anti-de Sitter. We parameterise the $AdS_{d+1}$ metric as
\begin{equation}
  \dd s^2 = \frac{\dd \rho^2}{4 \rho^2} + \frac{1}{\rho} \eta_{\mu \nu} \dd x^{\mu} \dd x^{\nu}
\end{equation}
where $\rho \to 0$ corresponds to the conformal boundary and $\eta_{\mu \nu}$ is the Minkowski metric. 

\begin{figure}
\begin{center}
\setlength{\unitlength}{0.50mm}
\includegraphics*[width=0.7\linewidth]{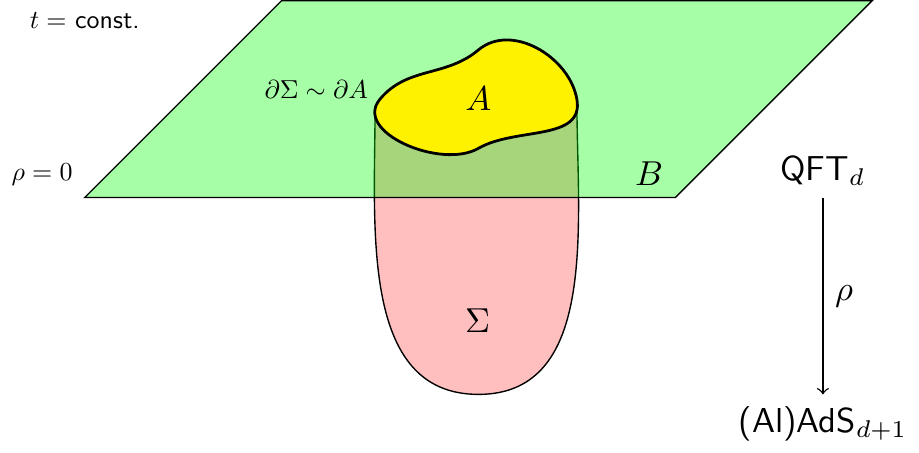}
\caption{The entangling surface embedded into the bulk manifold. }
\label{fig:surface1}
\end{center}
\end{figure}

The Ryu-Takayanagi function for the entanglement entropy is the area functional for a codimension two surface:
 \be
 S = \frac{1}{4 G_{d+1}} \int_{\Sigma} \dd^D \sigma^{\alpha} \, \sqrt{ \gamma} \label{rt}
\ee
where $G_{d+1}$ denotes the Newton constant (with the number of spatial dimensions in the field theory being $D = (d-1)$)  and $\gamma$ is the determinant of the induced metric on the surface. Throughout this section we work in a static setup, in which the entangling surface is independent of time. To find the bulk minimal surface $\Sigma$, we solve the equations of motion following from (\ref{rt}), subject to boundary conditions which define the entangling region in the dual field theory. In particular, as shown in Figure~\ref{fig:surface1}, the minimal surface $\Sigma$ has a conformal boundary $\partial \Sigma$ as $\rho \rightarrow 0$ which is conformal to the boundary $\partial A$ of the entangling region $A$ in the dual field theory.

\begin{figure}
\begin{center}
\setlength{\unitlength}{0.50mm}
\includegraphics*[width=0.7\linewidth]{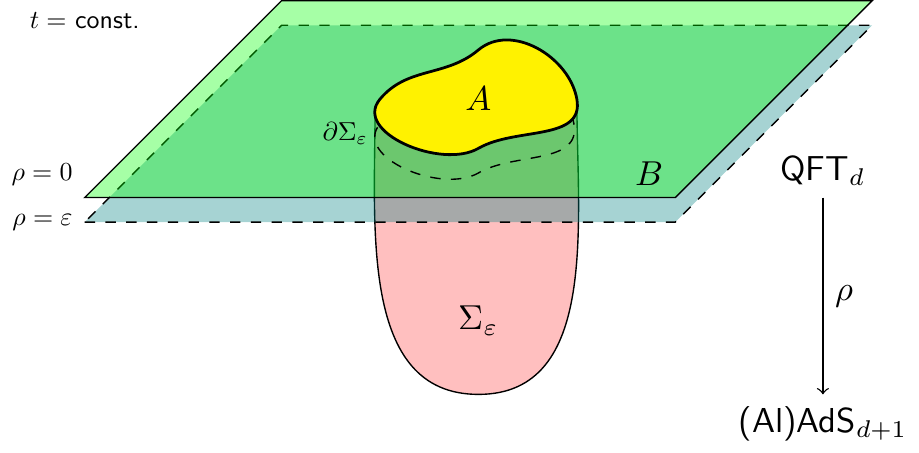}
\caption{The cutoff entangling surface.}
\label{fig:surface2}
\end{center}
\end{figure}

When one evaluates the onshell value of the functional (\ref{rt}), it has area divergences which may conveniently be regulated by setting $\rho = \epsilon$, see Figure~\ref{fig:surface2}. Let us denote the bulk manifold as ${\cal M}$ and the regulated
conformal boundary at $\rho = \epsilon$ as $\partial {\cal M}_{\epsilon}$. 
Since the entangling surface itself is asymptotically locally hyperbolic, the regulated functional (\ref{rt}) diverges as 
\be
S_{\rm reg} \sim \frac{{\cal A}_{\partial A}}{ \epsilon^{\frac{d}{2} -1}}  +\cdots
\ee
where ${\cal A}_{\partial A}$ is the area of the $(d-2)$-dimensional boundary of the entangling region $\partial A$. 

Following the principles of \cite{Henningson:1998gx,Henningson:1999xi,DeHaro2001}
we can now define a renormalized functional $S_{\rm ren}$ as
\be
S_{\rm ren} = {\cal L}_{\epsilon \rightarrow 0} \left ( S_{\rm reg} + S_{ct} \right )
\ee
where the counterterm action $S_{ct}$ is defined in terms of covariant properties of the boundary of the minimal surface and of the cutoff surface. Let the induced metric on the cutoff surface be $h_{\mu \nu}$ and the metric on the boundary of the minimal surface be $\tilde{\gamma}_{ab}$. Let us further denote the Ricci scalar of the boundary of the minimal surface as ${\cal R}$, with the corresponding Ricci tensor being ${\cal R}_{ab}$. Similarly we denote 
the extrinsic curvature of the minimal surface embedded into the cutoff surface as ${\cal K}_{ab}$ with trace ${\cal K}$ . Then counterterms must be expressible as 
\begin{equation}
S_{\rm ct} = \int_{\partial \Sigma} \dd^{D-1}x \sqrt{\tilde{\gamma}} {\cal L} \left ( {\cal K}, {\cal R}, {\cal R}_{ab} {\cal R}^{ab}, {\cal K}_{ab} {\cal K}^{ab}, \cdots  \right ),
\end{equation}
i.e. as a functional of extrinsic and intrinsic curvature invariants. In our setup there are three extrinsic curvatures arising from the following three different embeddings: the embedding of $\Sigma_\varepsilon$ in $\mathcal{M}_\varepsilon$, the embedding of $\del \Sigma_\varepsilon$ in $\Sigma_\varepsilon$, and the embedding of $\del \Sigma_\varepsilon$ in $\del \mathcal{M}_\varepsilon$. We should emphasise that it is the final one which is relevant for the counterterms, as the first two are not intrinsic to the regulated boundary. 

There is a further restriction on the allowed counterterms. The entanglement entropy of region $A$ is the same as the entanglement entropy of the complementary region $B$. If we require that the renormalized entanglement entropy satisfies the same property, then the counterterms should only depend on even powers of the extrinsic curvature ${\cal K}$, since the extrinsic curvature of $A$ is minus the extrinsic curvature of the complementary region $B$. 

Finally, we should note that the intrinsic and extrinsic curvature are related by Gauss-Codazzi relations. Throughout this section we will be interested in the case in which the background for the dual field theory is flat, in which case
\be
{\cal R} = {\cal K}^2 - {\cal K}_{ab} {\cal K}^{ab},
\ee
with analogous Gauss-Codazzi relations holding between higher order scalar invariants of the intrinsic and extrinsic curvature. 

\subsection{Explicit computation of counterterms}

Let us now express the area functional as
\begin{equation}
  S = \frac{1}{4 G_{d+1}} \int \dd \rho \int \, \dd^{D-1} \sigma^a \, \sqrt{ \gamma}
\end{equation}
where $\gamma_{\alpha \beta } = g_{mn} \partial_{\alpha} x^m  \partial_{\beta} x^n$, $g_{mn}$ is the metric on the $AdS$ target space and $x^m(x^{\alpha})$ defines the embedding in terms of the worldvolume coordinates $\sigma^{\alpha}$. We have implicitly fixed a static gauge, in which the time coordinate $t$ is constant and $\rho$ is one of the worldvolume coordinates, i.e. $\sigma^{\alpha} = \{ \rho, \sigma^a \}$. The spatial coordinates $x^i$ are then functions of $\rho$ and $\sigma^a$ and $D$ represents the number of spatial directions in the boundary theory.

In such a gauge the induced metric on the minimal surface is 
\begin{align}
  \gamma_{\rho \rho}&= \frac{1}{4 \rho^2} + \frac{1}{\rho} x^i_{,\rho} x^i_{,\rho} \\
\gamma_{\rho a}   &= \frac{1}{\rho} x^i_{,\rho} x^i_{,a} \qquad
\gamma_{a b}       = \frac{1}{\rho} x^i_{,a} x^i_{,b} \nonumber
\end{align}
where we denote $x^i_{,\rho} = \del_\rho x^i$ and $x^i_{,a} = \del_{\sigma^a} x^i$. One can often (but not always) further gauge fix, setting $x^a = \sigma^a$ and $x^{D} \equiv y (\rho, x^a)$, so that 
\begin{align}
  \gamma_{\rho \rho}&= \frac{1}{4 \rho^2} + \frac{1}{\rho} y_{,\rho} y_{,\rho} \\
\gamma_{\rho a}   &= \frac{1}{\rho} y_{,\rho} y_{,a} \qquad
\gamma_{a b}       = \frac{1}{\rho} (\delta_{ab} + y_{,a} y_{,b}),
\nonumber
\end{align}
reflecting the fact that a codimension one spatial minimal surface has only one transverse direction. Note however that such gauge fixing cannot be used to describe minimal surfaces with cusps, but in this paper we restrict to the case of surfaces without cusps.  

The gauge fixed minimal surface action is  given by
\begin{align}
  S &= \frac{1}{4 G_{d+1} } \int \dd \rho \int \dd^{D-1} x  \, \left( \frac{1}{4\rho^{D+1}}(1 + y_{,a} y_{,a}  + 4 \rho  y_{,\rho}^2) \right)^{1/2} \nonumber \\
  &= \frac{1}{4 G_{d+1}} \int \dd \rho \int \dd^{D-1} x \, \frac{1}{2 \rho^{(D+1)/2}} m(\rho, x^a)
\end{align}
where we have introduced the shorthand $m(\rho, x^a) = \sqrt{1 + y_{,a} y_{,a} + 4 \rho  y_{,\rho}^2}$. 

The regulated action is then of the form 
\begin{align}
  S_{\rm reg} &= \frac{1}{4 G_{d+1}} \int_{\epsilon} \dd \rho \int \dd^{D-1} x  \, \left( \frac{1}{4\rho^{D+1}}(1 + y_{,a} y_{,a}  + 4 \rho  y_{,\rho}^2) \right)^{1/2}  \label{regulated} \\
  &=\frac{1}{4 G_{d+1}} \int_{\partial \Sigma} \dd^{D-1} x  \sum \epsilon^{-k} (a_{k} (x) + \log \epsilon  b_k(x) ) + \cdots \nonumber
  \end{align}
where the explicit powers arising in the divergences and their coefficients $(a_k(x),b_k(x))$ are determined by analysing solutions to the minimal surface equations with the required boundary conditions asymptotically near the conformal boundary.

\bigskip

Note that the action does not depend explicitly on $y$ and the minimal surface equation is:
\begin{align}
  0 &= \partial_a \left( \frac{y_{,a}}{m \, \rho^{3/2}} \right) +  \partial_{\rho} \left( \frac{y_{, \rho}}{m \, \rho^{1/2}}\right), 
\end{align}
which should be solved near $\rho = 0$ subject to the boundary condition 
\begin{equation}
{\cal L}_{\rho \rightarrow 0} \left  (y(\rho,x^a)  \right ) = y_{(0)} (x^a),
\end{equation}
where $y_{(0)}(x^a)$ specifies the entangling region in the dual geometry. 

We wish to solve this equation iteratively for $y(\rho, x^a)$ as a series expansion in $\rho$. We consider the following Taylor series expansions for  $y(\rho, x^a)$:
\begin{align}
  y(\rho, x^a)  &= y_{(0)}(x) + y_{(\beta_1)}(x) \rho^{\beta_1} + y_{(\beta_2)}(x) \rho^{\beta_2} + \ldots
\end{align}
where we assume that  $0 < \beta_1 < \beta _2 < \ldots$. To solve the PDE we insert these expansions into the minimal surface equation and set $\rho = 0$. We then fix $\beta_1$ and $y_{(\beta_1)}$ to solve the resulting equation such that $y_{(0)}$ remains unconstrained. We then differentiate the minimal surface equation with respect to $\rho$ and repeat to find $\beta_2$.

After substituting the expansions into the minimal surface equation, one finds that the leading order terms are  $\rho^0$ and $\rho^{\beta_1 - 1}$. To leave $y_{(0)}$ unconstrained we must therefore set $\beta_1 = 1$ and deduce that:
\begin{equation}
  y_{(1)}(x) = 2 \sqrt{1 + y_{(0),a} y_{(0),a}} \partial_a \left ( \frac{ \partial_a y_{(0)}}{\sqrt{1+ y_{(0),a} y_{(0),a}}} \right ).
\end{equation}
To find higher terms in the asymptotic expansion we can use radial derivatives of the minimal surface equations. Before carrying out this procedure, let us consider the regulated onshell action \eqref{regulated} and determine the leading divergences, which are
\begin{equation}
S_{\rm reg} = \frac{1}{4 (D-1) G_{d+1} \epsilon^{\frac{1}{2}(D-1)} } \int_{\partial \Sigma} \dd^{D-1} x  \sqrt{1+ y_{(0),a} y_{(0),a}} + {\cal O} \left ( \epsilon^{\frac{3-D}{2}} \right )
\end{equation}
for $D > 1$. As anticipated above, this divergence scales with the area ${\cal A}_{\partial A}$ of the boundary of the entangling region 
\be
{\cal A}_{\partial A} = \int_{\partial \Sigma} \dd^{D-1} x  \sqrt{1+ y_{(0),a} y_{(0),a}}.
\ee
The case of $D=1$, corresponding to a dual two-dimensional conformal field theory, is degenerate. The divergence is logarithmic:
\begin{equation}
S_{\rm reg} = \frac{1}{8 G_3}  \Sigma_k y_k \log \epsilon
\end{equation}
with $y_k$ being the endpoints of the intervals defining the entangling region. The required counterterm action is therefore
\be
S_{\rm ct } =  - \frac{1}{8 G_3}  \Sigma_k y_k \log \left ( \frac{\epsilon}{\mu} \right ),
\ee
where $\mu$ is an arbitrary renormalization scale. 

\subsection{Entangling surfaces in $AdS_4$}

For minimal surfaces in $AdS_4$ the only divergence in the onshell functional is 
\begin{equation}
S_{\rm reg} = \frac{1}{4 G_4} \int_{\partial \Sigma} \dd x  \left ( \frac{1}{\epsilon^{\frac{1}{2}}}  \sqrt{1+ y_{(0), x} y_{(0),x}}  \right ),
\end{equation}
where the entangling region in the boundary is defined by a curve $y_{(0)}(x)$ in two dimensional space. Noting that the induced line element on the boundary of the entangling surface is 
\begin{equation}
\gamma^h_{xx} = \frac{1}{\epsilon} (1 + y_{,x} y_{,x} )
\end{equation}
the divergence is manifestly removed by the covariant counterterm
\begin{equation}
S_{\rm ct} = - \frac{1}{4 G_4} \int_{\partial \Sigma} \dd x \sqrt{\tilde{\gamma}},
\end{equation}
where $\tilde{\gamma}$ is the determinant of the induced metric on $\partial \Sigma$. 
This is the only possible divergent counterterm but the following counterterm is finite:
\begin{equation}
S_{\rm ct} = \frac{a_s}{4 } \int_{\partial \Sigma} \dd x \sqrt{\tilde{\gamma}} {\cal K}
\end{equation}
where ${\cal K}$ denotes the trace of the extrinsic curvature of the boundary of the minimal surface embedded into the regulated cutoff surface. For a curve $y(x,\epsilon)$ embedded into the cutoff surface
\begin{equation}
ds^2 = \frac{1}{\epsilon} \left ( -dt^2 + dx^2 + dy^2 \right )
\end{equation}
the trace of the extrinsic curvature is 
\begin{equation}
{\cal K} = \epsilon^{\frac{1}{2}} \frac{y_{,xx}}{(1+ y_{,x}^2)^{\frac{3}{2}}}
\end{equation}
and thus
\begin{equation}
\sqrt{\gamma^h} {\cal K} = \frac{y_{,xx}}{(1+ y_{,x}^2)} = \frac{y_{(0),xx}}{(1 + y_{(0),x}^2)} + {\cal O}(\epsilon),
\end{equation}
which is indeed finite.

Thus the complete renormalized action for the minimal surface is 
\begin{equation}
S_{\rm ren} = \frac{1}{4 G_4} \int_{\Sigma} d^2 \sigma \sqrt{\gamma} + \frac{1}{4 G_4} \int_{\partial \Sigma} dx \sqrt{\tilde{\gamma}} \left ( a_s {\cal K} - 1 \right ). 
\end{equation}
Note that terms depending on higher powers of the extrinsic curvature cannot contribute in the limit $\epsilon \rightarrow 0$. The finite counterterm is however not
consistent with the requirement that the renormalized entropy for any region is equal to that of its complement, and we must therefore set $a_s = 0$. 

\bigskip

As an example, let us evaluate the renormalized action for a disk entangling region, of radius $R$. The exact solution for the minimal surface is conveniently expressed 
in terms of the following coordinates
\begin{equation}
ds^2 = \frac{d \rho^2}{4 \rho^2} + \frac{1}{\rho} \left ( -dt^2 + dr^2 + r^2 d \phi^2 \right )
\end{equation}
as the circularly symmetric surface at constant time:
\begin{equation}
r^2 + \rho = R^2.
\end{equation}
The renormalized action for this surface is then 
\begin{equation}
S_{\rm ren} = - \frac{\pi}{2G_4}. 
\end{equation}
Note that this is independent of the choice of the radius $R$. 
Implicitly our Newton constant has been fixed to be dimensionless, as we chose the anti-de Sitter metric to have unit radius, absorbing the curvature radius into the overall prefactor of the bulk action. 
To reinsert the AdS radius we need only rescale the bulk metric by $\ell$ and the covariant counterterm by a further $\ell$. The result of these insertions is to simply rescale the results for the entanglement entropy by $\ell^2$:
\begin{align}
  S_{\rm ren} &=  - \frac{\pi \ell^2 }{2 G_N}, \label{disc}
\end{align}
where the Newton constant $G_N$ now has the standard dimensions. 
Since the dual field theory is conformal there is no other scale apart from $R$ and therefore $S_{\rm ren}$, which is dimensionless, cannot depend explicitly on $R$. 

\bigskip

Next we consider an entangling surface of two infinitely long parallel lines with separation $R$. We will regulate the lines to have length $L$ and by symmetry we may choose these lines to lie in the $x$ direction and to be located at $y = \pm \frac{R}{2}$. 
The minimal surface can be characterized by worldsheet coordinates $(\rho,x)$ and by symmetry the transverse coordinate $y$ depends only on $\rho$. The surface equations can be solved to obtain
\begin{equation}
  y(\rho) = \pm\left(- \frac{R}{2} + \frac{\rho^{3/2}}{3 \rho_0} {}_2F_1\left(\frac{1}{2}, \frac{3}{4}; \frac{7}{4}, \frac{\rho^2}{\rho_0^2} \right) \right).
\end{equation}
We can also rewrite this hypergeometric function in terms of the incomplete beta function $B_z(a,b)$ using the identity
\begin{equation}
  {}_2F_1(a,b;1+b;z) = b z^{-b} B_{z}(b,1-a).
\end{equation}
The surface has a turning point at $\rho_0$, where by symmetry $y(\rho_0) = 0$, and hence
\begin{equation}
  y(\rho_0) = 0 \implies \rho_0 = \frac{9\Gamma(5/4)^2}{4 \pi \Gamma(7/4)^2} R^2.
\end{equation}
The regularised holographic entanglement entropy is then given by
\begin{equation}
  S_{\rm reg} = \frac{L}{8 G_4} \int_\varepsilon^{\rho_0} \dd \rho \frac{\rho_0}{\sqrt{\rho^3(\rho_0^2 - \rho^2)}},
\end{equation}
This integral is elliptic and can be calculated analytically using
\begin{eqnarray}
  \int \frac{1}{\sqrt{w^3(a^2 - w^2)}} \dd w &=&  2 \frac{2 }{a^2 \sqrt{w}} \sqrt{a^2 - w^2}   \\ 
  & &  +  \frac{2}{\sqrt{a^3}} \left( F \left(\left.\sin^{-1}\left(\sqrt{\frac{w}{a}}\right)\right | -1\right)
   -E\left(\left.\sin^{-1}\left(\sqrt{\frac{w}{a}}\right)\right | -1\right) \right) \nonumber
\end{eqnarray}
where $F(\phi | k^2)$ and $E(\phi | k^2)$ are the incomplete elliptic integrals of the first and second kind respectively. 

The renormalized holographic entanglement entropy is:
\begin{equation}
  S_{\rm ren} = - \frac{\sqrt{2}\pi^2 \Gamma(7/4)}{3 G_4 \Gamma(1/4)^2 \Gamma(5/4)} \frac{L}{R}
\end{equation}
Note that in the above calculation we have implicitly assumed that $L \gg R$ and that there are no contributions from the lines $x = \pm L/2$, $-R/2 \le y \le R/2$. To take the limit of $L \rightarrow \infty$ we can calculate the renormalized entropy density 
\begin{equation}
s_{\rm ren} = {\cal L}_{L \rightarrow \infty } \left ( \frac{S_{\rm ren}}{L} \right ) = - \frac{\sqrt{2}\pi^2 \Gamma(7/4)}{3 R G_4 \Gamma(1/4)^2 \Gamma(5/4)}. 
\end{equation}

\bigskip

Finally let us consider the half plane entangling region with a boundary at $y=0$; again we regulate the $x$ direction to have length $L$. The bulk minimal surface has worldsheet coordinates $(\rho,x)$ and by symmetry $y=0$ over the surface. The regularised holographic entanglement entropy is
\begin{equation}
S_{\rm reg} = \frac{L}{ 8 G_4} \int^{\infty}_{\epsilon} \frac{d \rho}{\rho^{\frac{3}{2}}} = \frac{L}{4 G_4 \epsilon^{\frac{1}{2}}}
\end{equation}
and this term exactly cancels the counterterm giving
\begin{equation}
S_{\rm ren} = 0 \label{hp}
\end{equation}
which was to be expected since there is no other scale in the problem but $L$ and the dual theory is conformal. 

\bigskip

The calculation of the renormalized area of a two-dimensional minimal surface in four bulk dimensions has arisen in other contexts, including Wilson loops. In particular, 
anomalies were discussed in \cite{Henningson:1999xi} while 
the counterterm involving the regulated length of the boundary of the surface was discussed in the context of Wilson loops in \cite{Drukker:1999zq}; the counterterm was derived 
by requiring a well-defined variational principle. The relation of holographic renormalization to variational principles for minimal surfaces was discussed in detail in 
\cite{Papadimitriou:2010as}. 

Minimal surfaces in hyperbolic spaces were also analysed in \cite{Graham:1999pm}: generalizing  \cite{Henningson:1999xi}, it was noted that submanifold observables have conformal anomalies for specific codimensions. In particular, the results of  \cite{Graham:1999pm} imply that codimension two minimal surfaces in odd bulk dimensions have logarithmic divergences in their regulated volumes. This is consistent with our $D=1$ result above, and the $D = 3$ result we will give below. 

According to the results of \cite{Graham:1999pm} the renormalized area of a codimension two minimal surface in an even dimensional hyperbolic space should be a conformal invariant. This is not however apparent from the above results: the renormalized entropy of the half plane was found to be zero \eqref{hp}, while the renormalized entropy of the disk is finite and negative. Yet, as is well known, one can find a conformal bijective map between the disk and the half plane and therefore these entangling regions are conformally equivalent. We will explore this issue further in the next section. 

The renormalized entropy for the strip entangling region is negative. This is unsurprising: in \cite{Casini:2005zv,Casini:2006hu} the entanglement entropy for free scalars and  fermions was calculated for strip entangling regions and it was found that the entanglement entropy contains finite terms of the form
\begin{equation}
S_{\rm finite} = - k \frac{L}{R} 
\end{equation}
where again $L$ is the regulated length of the strip, $R$ is its width (with $L/R \gg1$) and $k$ is a {\em positive} constant, which takes the value of $k = 0.039$ for a real scalar and $k = 0.072$ for a Dirac fermion. 

\subsubsection{Relation to F theorem}

More generally, we should be unsurprised about finding negative values for the renormalized entanglement entropy. 
The conjectured F-theorem in three dimensions is the following. For a three-dimensional CFT we define the F quantity in terms of the (renormalized) partition function of the theory on a three sphere \cite{Jafferis:2011zi}, i.e. 
\begin{equation}
F = - \ln Z_{S^3} \label{f}
\end{equation}
and then the F theorem states that $F_{UV} \ge F_{IR}$. More precisely, in \cite{Jafferis:2011zi} it was conjectured that $F$ is positive in a unitary CFT, that it decreases along any RG flow  and that it is stationary at fixed points. Support for the conjecture can be found in \cite{Jafferis:2011zi,Klebanov:2011gs,Klebanov:2011td} and many subsequent works.

In odd spacetime dimensions
the finite terms in the entanglement entropy of a spherical region
\begin{equation}
S_{\rm finite} = (-)^{\frac{1}{2}(d-1)} 2 \pi a_{d} \label{ath}
\end{equation}
are conjectured to satisfy the relation $(a_d)_{UV} \ge (a_d)_{IR}$ for any RG flows between fixed points \cite{Myers:2010xs}. 
Indeed it has been shown that the sphere partition function and the sphere entanglement entropy are proportional
 using the CHM map \cite{Casini:2011kv}, thus establishing a connection between the F theorem and monotonous running of the finite part of the disk entanglement entropy. In three
 dimensions
 \be
 F = - 2 \pi a_3 
 \ee
 and hence positivity of F is equivalent to negativity of the finite parts of the entanglement entropy. 
 
To understand the relation between \eqref{f} and \eqref{ath} it is useful to recall the arguments of CHM \cite{Casini:2011kv} in more detail. Let us parameterise the flat three-dimensional metric in as
\begin{equation}
ds^2 = - dt^2 + dr^2 +r^2 d \phi^2
\end{equation}
Now consider the following change of coordinates
\begin{eqnarray}
t &=& R \frac{ \cos \theta \sinh \tau/R}{(1 + \cos \theta \cosh \tau/R )}; \\
r &=& R \frac{\sin \theta}{(1 + \cos \theta \cosh \tau/R )}; \nonumber
\end{eqnarray}
so that the metric becomes 
\begin{equation}
ds^2 = \Omega^2 \left ( - \cos^2 \theta d \tau^2 + R^2(d \theta^2 + \sin^2 \theta d \phi^2) \right )
\end{equation}
with conformal factor 
\begin{equation}
\Omega = (1 + \cos \theta \cosh \tau/R)^{-1}.
\end{equation}
One can clearly absorb the $R$ dependence as an overall factor by introducing $\tilde{\tau}= \tau/R$, so that the metric is conformal to the static patch of de Sitter space. Since $0 \le \theta < \pi/2$ the new coordinates cover $0 < r < R$, i.e. the disk of radius $R$ in the original flat coordinates, with $\theta \rightarrow \pi/2$ (the cosmological horizon) corresponding to $r = R$. The limits $\tau \rightarrow \pm \infty$ correspond to $t \rightarrow \pm R$ and therefore the new coordinates cover the causal development of the disk $r \le R$ from $t=0$. 

Modular transformations inside the causal development act as time translations in de Sitter space, and therefore the state in the de Sitter geometry is thermal with $\beta = 2 \pi R$. 
One can then identify the entanglement entropy for the disc of radius $R$ in flat space with the thermodynamic entropy of the thermal state in de Sitter space, which in turn is given by 
\begin{equation}
S_{\rm de Sitter} = - W
\end{equation}
where $W = - \ln Z$ is the free energy of the partition function $Z$. This relation is the origin of the above statement that the disc entanglement entropy is related to the partition function on the sphere, since the analytic continuation of de Sitter is the three-dimensional sphere. 

The corresponding Euclidean transformations begin from the metric
\begin{equation}
ds^2 = dt_{E}^2 + dr^2 +r^2 d \phi^2
\end{equation}
with the transformations being
\begin{eqnarray}
t_{E} &=& R \frac{ \cos \theta \sin \tau_E/R}{(1 + \cos \theta \cos \tau_E/R )}; \\
r &=& R \frac{\sin \theta}{(1 + \cos \theta \cos \tau_E/R )}; \nonumber
\end{eqnarray}
so that the metric becomes 
\begin{equation}
ds^2 = \Omega^2 \left ( \cos^2 \theta d \tau_E^2 + R^2(d \theta^2 + \sin^2 \theta d \phi^2) \right )
\end{equation}
with conformal factor 
\begin{equation}
\Omega = (1 + \cos \theta \cos \tau_E/R)^{-1}. \label{chm5}
\end{equation}
In the transformed coordinates the Euclidean time $\tau_E$ is periodic with period $2 \pi R$ for the sphere to be regular and $0 \le \theta < \pi/2$. 
 
 Implicitly the finite parts of the partition function on the $S^3$ are computed by renormalization; the CHM map thus relates the (renormalized) F quantity to the 
 the corresponding renormalized entanglement entropy i.e.
 \be
 F = - S_{\rm ren}
 \ee
 with $F$ being positive and decreasing along an RG flow. For the disk entangling region we thus find holographically that 
 \be
 F = \frac{\pi}{2 G_4} \label{disc3}
 \ee
which is indeed positive. 
 
\bigskip

Let us now review the evaluation of the partition function on $S^3$ for a conformal field theory with a holographic dual described by Einstein gravity. The renormalized partition function is then calculated by evaluating the renormalized Euclidean action \cite{DeHaro2001}:
\begin{equation}
I = - \frac{1}{16 \pi G_4} \int d^4 x \sqrt{g} ( R_g + 6) + \frac{1}{8 \pi G_4} \int d^3 x \sqrt{h} (1 - \frac{ R}{4} ), \label{Action}
\end{equation}
where $R_g$ is the bulk Ricci scalar and $R$ is the Ricci scalar for the boundary metric $h$. 
For the $AdS_4$ geometry with spherical slicing
\begin{equation}
ds^2 = d \rho^2 + \sinh^2 \rho d \Omega_3^2
\end{equation}
the renormalized onshell action is then 
\begin{equation}
I = \frac{\pi}{2 G_4}. \label{onshell}
\end{equation}
Comparing \eqref{onshell} with \eqref{disc3}, the values indeed agree.
Note that there is no ambiguity in the holographically renormalized action (\ref{Action}): there are no candidate covariant finite counterterms. We will explain further in section \ref{five} how the renormalization schemes for the bulk action and for the entanglement entropy are related.

\subsection{Renormalization for AdS in general dimensions}

In this section we describe the holographic renormalization of the  entanglement entropy for AdS in general dimensions, noting the generic forms of possible counterterms, anomalies, and finding the first two counterterms.

We begin by establishing the notational conventions we will use in this section. We will take our bulk manifold $\mathcal{M}$ to be $AdS_{D+2}$ and will work exclusively in coordinates in which the metric takes the form
\begin{equation}
  \dd s^2_{\mathcal{M}} = g_{mn} \dd x^m \dd x^n = \frac{\dd \rho^2}{4 \rho^2} - \frac{1}{\rho} \dd t^2 + \frac{1}{\rho}\delta_{ij}\dd x^i \dd x^j
\end{equation}
where $i,j = 1, \ldots, D$ are the boundary spatial directions. The entangling surface $\Sigma$ is a codimension 2 surface of $\mathcal{M}$ satisfying the appropriate boundary conditions. We choose the coordinates on $\Sigma$ to be $(\rho, x^a)$ where $a = 1, \ldots, D-1$. The embedding of $\Sigma$ in $\mathcal{M}$ is then given by:
\begin{equation}
  X^m = (\rho, t, x^1, \ldots, x^{D-1}, y(\rho, x^a))
\end{equation}
where $t$ is a constant. This is an appropriate gauge whenever the boundary entangling region specified by $y(0, x^a)$ is smooth.

We regulate the bulk as $\mathcal{M}_\varepsilon$ by restricting $\rho \geq \varepsilon > 0$, and similarly define the regulated entangling surface $\Sigma_\varepsilon$ by the same restriction. The surface $\Sigma_\varepsilon$ is a  constant time hypersurface of $\mathcal{M}_\varepsilon$.
The metric $\gamma_{\alpha \beta}$ on $\Sigma_\varepsilon$ is given by
\begin{equation}
  \dd s^2_{\Sigma_{\varepsilon}} = \left(\frac{1}{4\rho^2} + \frac{1}{\rho} y_{,\rho}^2\right)\dd \rho^2 + \frac{2}{\rho} y_{,\rho} y_{,a} \dd x^a +  \frac{1}{\rho}(\delta_{ab} + y_{,a} y_{,b}) \dd x^a \dd x^b.
\end{equation}
In this gauge the regulated bare entanglement entropy is given by
\begin{equation}
  S_{\rm reg} = \frac{1}{4 G_{D+2}} \int_{\del \Sigma_\varepsilon} \dd^{D-1} \int_{\varepsilon}^{\rho_0} \dd \rho \frac{1}{2\rho^{(D+1)/2}} \sqrt{1 + 4 \rho y_{,\rho}^2 + y_{,a}^2}
\end{equation}
where summation is implicit for the $a,b,\ldots$ indices. 

From the action one can find the equation of motion for $y(\rho, x^a)$, as in the previous section. Expanding the solution near the conformal boundary one finds:
\begin{equation}
  y(\rho, x^a) = y^{(0)} + y^{(1)}\rho + O(\rho^2); \qquad y^{(1)} = \frac{1}{2(D-1)}\left(y^{(0)}_{,aa} - \frac{y^{(0)}_{,a} y^{(0)}_{,ab} y^{(0)}_{,b}}{1 + {y^{(0)}_{,c}}^2}\right).
\end{equation}
Note that this result agrees with the $D = 2$ case we considered above. 

Inserting the asymptotic expansion into the regulated functional  yields for $AdS_5$ ($D=3$)
\begin{equation}
  S_{\rm reg} = \frac{1}{4 G_{5}}\int_{\del \Sigma_\varepsilon}\dd^{2 }x (1 + {y^{(0)}_{,c}}^2)^{1/2} \left( \frac{1}{2 \varepsilon} - \frac{y^{(0)}_{,a} y^{(1)}_{,a} + 2 {y^{(1)}}^2}{2 (1 + {y^{(0)}_{,b}}^2)} \ln \varepsilon + \ldots \right)
\end{equation}
where the ellipses denote finite terms. Similarly for $D > 3$
\begin{equation}
  S_{\rm reg} = \frac{1}{4 G_{D+2}}\int_{\del \Sigma_\varepsilon}\dd^{D-1}x (1 + {y^{(0)}_{,c}}^2)^{1/2} \left( \frac{\varepsilon^{-\frac{(D-1)}{2}}}{D-1} + \frac{\varepsilon^{-\frac{(D-3)}{2}}}{D-3} \frac{y^{(0)}_{,a} y^{(1)}_{,a} + 2 {y^{(1)}}^2}{1 + {y^{(0)}_{,b}}^2} + \ldots \right)
\end{equation}
where the ellipses denote subleading divergences and terms that are finite as $\varepsilon \to 0$.


Our task is now to find counterterms which are integrals of covariant quantities defined on $\del \Sigma_\varepsilon$, i.e. scalars constructed from the intrinsic and extrinsic curvature tensors. The induced metric on $\del \Sigma_\varepsilon$, $\tilde{\gamma}_{ab}$ is given by
\begin{equation}
  \dd s^2_{\tilde{\gamma}} = \frac{1}{\varepsilon}(\delta_{ab} + y_{,a}y_{,b}) \dd x^a \dd x^b
\end{equation}
which has determinant
\begin{equation}
  \tilde{\gamma} = \det(\tilde{\gamma}_{ab}) = \varepsilon^{-\frac{D-1}{2}}(1 + y_{,a}^2)
\end{equation}
by Sylvester's determinant theorem. 

Using the asymptotic expansion we can expand the volume form to first subleading order in $\varepsilon$ as:
\begin{equation}
  \sqrt{\tilde{\gamma}} = \varepsilon^{-\frac{D-1}{2}}(1 + {y^{(0)}_{,c}}^2)^{1/2}\left(1 + \varepsilon \frac{y^{(0)}_{,b}y^{(0)}_{,b}}{1 + {y^{(0)}_{,c}}^2} + \ldots \right).
\end{equation}
On dimensional grounds we can show that all curvature scalars will be at least $O(\varepsilon^{1/2})$ and so we can uniquely identify the leading divergence in $S_{\rm ren}$ as coming from the area divergence, as expected. Our first counterterm is therefore
\begin{equation}
  S_{ct,1} = - \frac{1}{4G_{D+2}} \frac{1}{D-1} \int_{\del \Sigma_\varepsilon} \dd^{D-1} x \sqrt{\tilde{\gamma}}
\end{equation}
which is again consistent with our previously found $AdS_4$ ($D=2$) result.

We now need to find the counterterms for the  subleading divergences. Let us consider first the case of $D=3$.
Using integration by parts we can rewrite:
\begin{equation}
  \int_{\del \Sigma_\varepsilon} \dd^{D-1}x \frac{y_{,A}^{(0)}y_{,A}^{(1)}}{(1 + {y^{(0)}_{,C}}^2)^{1/2}} = - \int_{\del \Sigma_\varepsilon} \dd^{D-1}x \frac{{y^{(1)}}^2}{(1 + {y^{(0)}_{,C}}^2)^{1/2}}
\end{equation}
and hence for $D=3$
\be
S_{\rm reg} + S_{ct,1} =  - \frac{1}{8 G_5 } \int_{\del \Sigma_\varepsilon} \dd^{2}x \sqrt{\tilde{\gamma}} \frac{ {y^{(1)}}^2}{1 + {y_{,C}^{(0)}}^2} \ln \varepsilon + \ldots \label{D3}
\ee
To rewrite this term covariantly we note that the metric on a constant time hypersurface of the regulated boundary is given by
\begin{equation}
  \dd s^2_{D} = \tilde{g}_{ij} \dd x^i \dd x^j = \frac{1}{\varepsilon} \delta_{ij} \dd x^i \dd x^j
\end{equation}
and the embedding of $\del \Sigma_\varepsilon$ is given by $x^D \equiv y(\varepsilon, x^a)$. The unit normal covector is then given by
\begin{equation}
  n_\flat = \varepsilon^{-\frac{1}{2}}(1 + y_{,c}^2)^{- \frac{1}{2}}( y_{,a} \dd x^a - \dd x^D)
\end{equation}
From this we define the induced metric $\tilde{\gamma}_{ij}$, and the extrinsic curvature ${\cal K}_{ij}$ by
\be  
\tilde{\gamma}_{ij} = \tilde{g}_{ij} - n_i n_j \qquad
 {\cal K}_{ij} = \tilde{\gamma}^k_i \nabla_k n_j
\ee
where $\nabla_k$ is the covariant derivative with respect to $\tilde{g}_{ij}$. The trace of the extrinsic curvature is then given by
\be
  {\cal K} = \frac{\varepsilon^{\frac{1}{2}}}{(1 + y_{,c}^2)^{\frac{1}{2}}} \left( y_{,aa} - \frac{y_{,a}y_{,ab}y_{,b}}{1 + y_{,e}^2} \right) 
 = 2(D-1) \frac{\varepsilon^{\frac{1}{2}} y^{(1)}}{(1 + {y^{(0)}_{,c}}^2)^{\frac{1}{2}}} + \ldots \label{kD}
\ee
By contrast, the Ricci scalar  has a qualitatively different structure:
\be
{\cal R} =  \frac{2 \varepsilon}{(1 + y_{,e}^2)} \left( y^2_{,aa} - y_{,ab}y_{,ab} - y_{,a} y_{,b} \frac{y_{,ab} y_{,cc} - y_{,ac} y_{,bc}} {1 + y_{,d}^2} \right) \label{rsc1}
\ee
Comparing with (\ref{D3}) we can see that the required logarithmic counterterm is hence written in terms of the extrinsic curvature as 
\be
S_{ct,2} =  \frac{1}{64 G_5 } \int_{\del \Sigma_\varepsilon} \dd^{2}x \sqrt{\tilde{\gamma}}{\cal K}^2  \ln  \frac{\varepsilon}{\mu}, \label{anom2}
\ee
with $\mu$ a cutoff scale. 

Similarly for $D > 3$ we can show that 
\begin{equation}
  S_{EE} + S_{ct,1} = \frac{1}{4 G_{D+2}}\frac{2}{(D-3)} \int_{\del \Sigma_\varepsilon} \dd^{D-1}x \sqrt{\tilde{\gamma}} \frac{\varepsilon {y^{(1)}}^2}{1 + {y_{,C}^{(0)}}^2} + \ldots
\end{equation}

At this order the only possible intrinsic curvature term would be ${\cal R}$, the Ricci scalar on $\del \Sigma_\varepsilon$ but from \eqref{rsc1} this does not have the right structure to be the correct counterterm. Using (\ref{kD}) we can show that the required counterterm is
\begin{equation}
  S_{ct,2} = - \frac{1}{8 G_{D+2}} \frac{1}{(D-1)^2(D-3)} \int_{\del \Sigma_\varepsilon} \dd^{D-1} x \sqrt{\tilde{\gamma}} {\cal K}^2.
\end{equation}
Note that other extrinsic curvature invariants again either do not have the correct $\varepsilon$ structure or the correct ${y^{(1)}}^2$ behaviour to arise as possible counterterms.

In the $D=2$ analysis  we found that ${\cal K}$ would be a finite counterterm, but it is excluded by the requirement that the renormalised entanglement entropy of the complementary region is equal to that of the original region. For $D > 2$ we can further note that 
\begin{equation}
  \int_{\del \Sigma_\varepsilon} \dd^{D-1} x \sqrt{\tilde{\gamma}} {\cal K}^{D-1}
\end{equation}
are finite counterterms. The complementarity requirement rules out such counterterms for even $D$ (corresponding to field theories in odd spacetime dimensions). 
For odd $D$ (corresponding to field theories in even spacetime dimensions), these counterterms are consistent with the requirement that the entropy of the complement is the same as that of the original region. Indeed we already saw that such a term arises in $D=3$: it is automatically included in \eqref{anom2} in the $\mu$ dependent part. 

In addition, higher dimensions allows the possibility of other finite counterterms constructed from curvature invariants such as
\begin{equation}
 \int_{\del\Sigma_\varepsilon} \dd^{D-1} x \sqrt{\tilde{\gamma}} ({\cal K}_{ab} {\cal K}^{ab})^{(D-1)/2}; \quad \int_{\del\Sigma_\varepsilon} \dd^{D-1} x \sqrt{\tilde{\gamma}} \tilde{R}^{(D-1)/2}
\end{equation}
which are both valid for odd $D$, so that they are analytic. These counterterms are however not  linearly independent of each other, due to the Gauss-Codazzi relations.
In general there will always be finite counterterms possible in even spacetime dimensions and the number of such terms will increase with $D$ implying there are an increasing number of scheme dependent terms. We will understand in section \ref{five} how these finite counterterms relate to the scheme dependence of the partition function. 

\bigskip

Thus, to summarise the results in this section, the renormalized entanglement entropy for static surfaces in $AdS_{D+2}$ is 
\bea
S_{\rm ren} &=& \frac{1}{4 G_{D+2}} \int_{\Sigma} \dd^D \sigma^{\alpha} \, \sqrt{ \gamma}  \\ 
&& \qquad - \frac{1}{4G_{D+2}}  \int_{\del \Sigma} \dd^{D-1} x \sqrt{\tilde{\gamma}}
\left (  \frac{1}{D-1} + \frac{1}{2(D-1)^2(D-3)} {\cal K}^2 \cdots \right ), \nn
\eea
where $D$ represents the number of spatial dimensions in the dual field theory. The first counterterm is logarithmic for $D=1$. Only the first
counterterm given above is needed for gravity in four bulk dimensions ($D=2$). The second counterterm is logarithmic at $D=3$ and is needed in the form given above for $D > 3$.
Additional counterterms involving higher order curvature invariants are needed for $D \ge 5$; the additional counterterms are associated with logarithmic divergences (i.e. conformal anomalies) in odd dimensions.  

The analysis in this section assumed a Poincar\'{e} parameterisation of $AdS_{D+2}$, i.e. we assumed a flat background metric for the dual field theory.  We will generalize these results in section \ref{five}. Note that although the renormalized
entanglement entropy can be covariantized as shown in section \ref{five} the complete holographic dictionary would also need to take into account real time issues \cite{Skenderis2009} for non-static setups. 

\section{Entanglement entropy for holographic RG flows} \label{four}

A holographic RG flow (for a field theory in a flat background) can be described by a domain wall geometry
\be
\dd s^2 = d w^2 + e^{2 {\cal A}(w)} dx^{\mu} dx_{\mu}
\ee
where the warp factor ${\cal A}(w)$ is linear in $w$ at a fixed point. The geometry satisfies the equations of motion derived from Einstein gravity
coupled to scalar fields $\phi^A$, and the scalar fields have corresponding radial profiles $\phi^A(w)$. In what follows we will consider the case of a single scalar field
with the bulk  action being 
\begin{equation}
    I = \frac{1}{16 \pi G_4} \int \dd^4 x \, \sqrt{-g}\left( R_g - \frac{1}{2}{(\del \phi)}^2 + V(\phi)  \right),
\end{equation}
with $V(\phi)$ being the scalar potential. The generalisation to multiple scalar fields would be straightforward. 

We restrict to UV conformal theories, so that
the scalar potential $V(\phi)$ can be expanded as a power series in $\phi$ near the boundary:
\begin{equation}
  V(\phi) = 6 - \sum_{n=1}^\infty \frac{\lambda_{(2n)}}{(2n)!} \phi^{2n}.
\end{equation}
The mass $M$ of the scalar is then given by $M^2 = \lambda_{(2)}$, so the scalar field is dual to a dimension $\Delta$ operator in the boundary CFT where $M^2 = \Delta(\Delta-3)$. In what follows we will denote 
\be
\Delta_{+} = \frac{3}{2} + \frac{1}{2} \sqrt{9 + 4M^2}.
\ee
For $-9/4 < M^2 < - 5/4$, two quantizations are possible with the operator dimension corresponding to the second quantization being
\be
\Delta_{-} = \frac{3}{2} - \frac{1}{2} \sqrt{9 + 4M^2}.
\ee
The equations of motion are
\bea
&& \ddot{\cal A} = - \frac{1}{4} (\dot{\phi})^2 \\
&& \ddot{\phi} + 3 \dot{\cal A} \dot{\phi} = - \frac{d V}{d \phi}  \nonumber
\eea
where a dot denotes a derivative with respect to $w$. It is well-known, see \cite{Freedman:2003ax}, that these equations are always equivalent to first order equations 
\be
\dot{\cal A} =   W \qquad
\dot{\phi} = - 4  \frac{d W}{d {\phi}}
\ee
where the superpotential $W(\phi)$ is given by
\be
V = - 2 \left ( 4 \left ( \frac{d W}{d \phi} \right )^2 - 3 W^2 \right ), 
\ee
with
\be
W = 1 + \frac{1}{8} (3 - \Delta_{+}) \phi^2 + \cdots  
\ee
Note that the superpotential is not unique at higher orders in the scalar field: different choices are associated with different RG flows and in a supersymmetric theory only one choice will be supersymmetric. 
For flat sliced domain walls corresponding to holographic RG flows, the appropriate counterterm for the bulk action can be expressed in terms of the superpotential as
\be
I_{\rm ct} = - \frac{1}{4 \pi G_4} \int_{{\cal M}_{\epsilon}} \dd^3 x \sqrt{-h} W.
\ee
To match with conventions in earlier sections, it is convenient to express the asymptotically AdS$_4$ domain wall spacetime in the coordinates 
\begin{equation}
  \dd s^2 = \frac{\dd \rho^2}{4 \rho^2} + \frac{1}{\rho}e^{A(\rho)} \eta_{\mu \nu} \dd x^{\mu} \dd x^{\nu}
\end{equation}
where $\rho \to 0$ corresponds to the conformal boundary, $\eta_{\mu \nu}$ is the flat metric, with coordinates $(t, x, y)$. Near the conformal boundary 
\begin{equation}
  e^{A(\rho)} = 1 + \ldots,
\end{equation}
where the subleading terms depend on the form of the scalar potential. In these coordinates the Einstein and scalar equations become
\bea
&& A'' + \frac{1}{\rho} A' = - \frac{1}{4} (\phi')^2; \label{ein-fe} \\
&& 4 \rho^2 \phi'' + 2 (3 \rho A' - 1) \rho \phi' = - \frac{d V}{d \phi}. \nn
\eea
These equations can also be rewritten in terms of the superpotential as 
\be
-  \rho A' = \tilde{W} \qquad
\rho \phi' = 2 \frac{ d\tilde{W}}{d \phi}
\ee
where $\tilde{W} = W  -1$. 

\subsection{Renormalization of entanglement entropy}

Consider a codimension two minimal spacelike surface probing the domain wall spacetime. 
The entanglement entropy functional is
\begin{equation}
  S = \frac{1}{4 G_4}  \int \dd \rho \, \dd x \, \sqrt{ \gamma}
\end{equation}
where $\gamma_{\mu \nu} = g_{mn} \partial_{\mu} X^m(\rho, x) \partial_{\nu} X^n(\rho, x)$, $g_{mn}$ is the metric on the full target space, and $X^m(\rho, x)$ is the embedding. We will again work in static gauge where the embedding is given by $X^m(\rho,x) = (\rho, t, x, y(\rho, x))$ and $t$ is a constant. Therefore
\begin{align}
  \gamma_{\rho \rho}&= \frac{1}{4 \rho^2} + \frac{e^{A(\rho)}}{\rho}y_{\rho}^2
\\\gamma_{\rho x}   &= y_{\rho} y_{x} \frac{e^{A(\rho)}}{\rho} \nonumber
\\\gamma_{xx}       &= \frac{1}{\rho} e^{A(\rho)} + \frac{e^{A(\rho)}}{\rho}y_{x}^2 \nonumber
\end{align}
where we denote $y_{\rho} = \del_\rho y$ and $y_{x} = \del_x y$. The entanglement entropy is thus given by
\begin{align}
  S 
  &= \frac{1}{4 G_4} \int \dd \rho \, \dd x \, \frac{e^{A(\rho)/2}}{2 \rho^{3/2}} \sqrt{1 + y_{x}^2 + 4 \rho e^A y_{\rho}^2} \\
  &= \frac{1}{4 G_4} \int \dd \rho \, \dd x \, \frac{e^{A/2}}{2 \rho^{3/2}} m(\rho, x) \nonumber
\end{align}
where we have introduced the shorthand $m(\rho, x) = \sqrt{1 + y_{x}^2 + 4\rho e^{A} y_{\rho}^2}$.

The minimal surface equation is:
\begin{align}
  0 &= (1 + 4 \rho e^A y_\rho^2)y_{xx} + \rho e^A (1 + y_x^2) y_{\rho\rho} - 5 \rho e^A y_\rho y_x y_{\rho x} \\
   &+ \frac{1}{2} e^A y_\rho (\rho A'(3 + 3 y_x^2 + 8 \rho e^A y_\rho^2) - 1 - y_x^2 - 8 \rho e^A y_\rho^2). \nn
   \label{eq:min_surface_pde}
\end{align}
We now solve this equation iteratively for $y(\rho, x)$ as a series expansion in $\rho$. We assume the following Taylor series expansions for $A(\rho)$ and $y(\rho, x)$:
\begin{align}
  e^{A(\rho)} &= 1 + A_{(\alpha)} \rho^\alpha + \ldots \\
  y(\rho, x)  &= y_{(0)}(x) + y_{(\beta_1)}(x) \rho^{\beta_1} + y_{(\beta_2)}(x) \rho^{\beta_2} + \ldots \nn
\end{align}
where we assume that $\alpha > 0$ and $0 < \beta_1 < \beta _2 < \ldots$. To solve the PDE we insert these expansions into equation~\eqref{eq:min_surface_pde} and set $\rho = 0$. We then fix $\beta_1$ and $y_{(\beta_1)}$ to solve the resulting equation to leave $y_{(0)}$ unconstrained and differentiate equation~\eqref{eq:min_surface_pde} with respect to $\rho$ and repeat to find $\beta_2$.

After substituting the expansions into the minimal surface equation, one finds that the leading order behaviour is a term constant in $\rho$ and a term scaling as $\rho^{\beta_1 - 1}$. To leave $y_{(0)}$ unconstrained we must set $\beta_1 = 1$ (as before) and deduce that:
\begin{equation}
  y_{(1)}(x) = \frac{2 y_{(0) xx}}{1 + ({y_{(0)x}})^2}.
\end{equation}
Next we substitute the expansions into the $\rho$ derivative of equation~\eqref{eq:min_surface_pde}. In all cases the lowest power involving $\beta_2$ is $\rho^{\beta_2 - 2}$ and we choose $\beta_2$ so as to cancel the leading order divergence involving $\alpha$.

In the case that $\alpha < 1$ the leading order divergence involving $\alpha$ goes as $\rho^{\alpha - 1}$ which requires $\beta_2 = 1 + \alpha$ and the following value of $y_{(1 + \alpha)}$ to cancel the divergence:
\begin{equation}
  y_{(1+ \alpha)} = -\frac{A_{(\alpha)}(3\alpha - 1) y_{(1)}}{2 \alpha^2 + \alpha - 1}
\end{equation}
Note that the denominator here vanishes when $\alpha = \frac{1}{2}$ (and when $\alpha = -1$ which is excluded by the boundary conditions)  and this case needs to be treated separately. 

In the case $\alpha > 1$ the leading order term involving $\alpha$ is not divergent and we can set $\beta_2 = 2$ with
\begin{equation}
  y_{(2)} = \frac{4y_{(1)}^3 + 6 y_{(1)}y_{(0)x}y_{(1)x} - 4 y_{(1)}^2 y_{(0)xx} - y_{(1)xx}}{1 + ({y_{(0)x}})^2}
\end{equation}
In the case where $\alpha = 1$ these two results overlap to give $\beta_2 = 2$ and
\begin{equation}
  y_{(2)} = \frac{4y_{(1)}^3 + 6 y_{(1)}y_{(0)x} - 4 y_{(1)}^2 y_{(0)xx} - y_{(1)xx}}{1 + ({y_{(0)x}})^2} - A_{(1)} y_{(1)}.
\end{equation}
One can similarly analyse the asymptotic expansions to higher order but this will not be needed in calculating the regularised entanglement entropy.  

Let us now turn to the regularisation of the entanglement entropy functional. Using the series expansion for $y(\rho,x)$ the
small $\rho$ behaviour of the action is
\begin{equation}
  S =\frac{1}{4 G_4}  \int \dd x \, \dd \rho \frac{e^{A/2}}{2 \rho^{3/2}} \sqrt{1 + ({y_{(0)x}})^2} \left(1 + \frac{1}{2} \rho B(\rho, x) + \ldots \right).
\end{equation}
where $B(\rho, x)$ is a function which is constant in $\rho$ to leading order. The full expression for $B(\rho, x)$ is given by
\begin{equation}
  \rho B(\rho, x) = \frac{y_x^2 + 4 \rho e^A y_\rho^2 - ({y_{(0)x}})^2}{1 + ({y_{(0)x}})^2}
\end{equation}
where it is understood that the series expansions for $e^A$ and $y$ are inserted above. It is clear that
\begin{equation}
  \int \dd x \, \int_\varepsilon \dd \rho \frac{e^{A/2}}{4}\sqrt{1 + ({y_{(0)x}})^2} \rho^{-1/2} B \sim \varepsilon^{1/2} + \ldots
\end{equation}
which vanishes as the cutoff is removed.

Hence to find the regularised action we only need to expand the function $e^{A(\rho)/2}$ and keep terms which are powers of $\rho^{1/2}$ or lower:
\be
S_{\rm reg} =\frac{1}{4 G_4}  \int \dd x \, \sqrt{1 + ({y_{(0)x}})^2} \int \dd \rho \frac{e^{A/2}}{2 \rho^{3/2}} 
\ee
The latter radial integral depends only on the background and not on the specific embedding. 

The first counterterm we require is needed in all cases independently of $\alpha$: this is the volume divergence associated with the asymptotically AdS background. The necessary counterterm here is as before 
\begin{equation}
  S_{\rm ct} = - \frac{1}{4 G_4} \int_{\partial \Sigma} \dd x \, \sqrt{\tilde{\gamma}}.
\end{equation}
The remaining divergent terms depend explicitly on $A_{(\alpha)}$ and $\alpha$. These terms can only be non-trivial if there is a non-trivial matter content in the bulk and consequentially the counterterms must be functions of the scalar fields on the $\rho = \varepsilon$ slice pulled back on to the minimal surface. 

Solving the field equations \eqref{ein-fe} to leading orders in $\rho$ implies that 
\be
\phi = \phi_{(0)} \rho^{\frac{1}{2}(3 - \Delta_+)} + \cdots
\ee
and for the warp factor:
\begin{equation}
  \alpha = 3-\Delta_+; \qquad A_{(\alpha)} = - \frac{1}{8}\phi_{(0)}^2.
\end{equation}
Subleading divergences in the entanglement entropy are only present when $\Delta_+ > 5/2$. The regulated onshell action up to the first subleading divergence is:
\begin{equation}
  S_{\rm reg} = \frac{1}{4 G_4} \int_{\partial \Sigma_{\epsilon}} \dd x \, \sqrt{\tilde{\gamma}} \left( 1 + \frac{3-\Delta_+}{8(5-2\Delta_+)} \phi_{(0)}^2 \varepsilon^{3-\Delta_+} + \ldots \right)
\end{equation}
and to leading order we also know that on the $\rho = \varepsilon$ hypersurface  $\phi = \phi_{(0)} \varepsilon^{(3-\Delta_+)/2}$ so it is simple to write the $n=1$ divergence in a covariant form so that the corresponding counterterm can then be read off:
\begin{equation}
  S_{\rm ct} = - \frac{1}{4 G_4} \int_{\partial \Sigma} \dd x \, \sqrt{\tilde{\gamma}} \frac{3-\Delta_+}{8(5-2\Delta_+)} \phi^2.
\end{equation}
At $\Delta_+ = 5/2$ the divergence becomes logarithmic and is associated with a conformal anomaly; we will discuss such anomalies further below. 

\bigskip

Given a superpotential for the RG flow one can find an exact expression for the counterterms to all orders as follows. We have argued that the counterterms can be written covariantly as
\be
S_{\rm ct} = - \frac{1}{4 G_4} \int_{\partial \Sigma} \dd x \, \sqrt{\tilde{\gamma}} Y(\phi)
\ee
where $Y(\phi)$ is analytic in the scalar field. (Here we exclude conformal anomalies, which we will discuss below.)
By construction the counterterm is chosen to cancel divergences and hence
\be
\int_{\epsilon} d \rho \frac{e^{\frac{A}{2}}}{2 \rho^{\frac{3}{2}}} = \frac{e^{\frac{A}{2}}}{\epsilon^{\frac{1}{2}}} Y(\phi),
\ee
where implicitly the latter is evaluated at $\rho = \epsilon$. Differentiating this expression with respect to the radius we then obtain
\be
A' Y + 2 \frac{d Y}{d \phi} \phi' - \frac{Y}{ \rho} = - \frac{1}{\rho}.
\ee
One can then substitute in the superpotential to get
\be
(1 + \tilde{W})Y - 4 \frac{d Y}{d \phi} \frac{d \tilde{W}}{d \phi}  =  1, \label{counterp}
\ee
i.e. an expression for $Y(\phi)$ in terms of the superpotential $\tilde{W}(\phi)$ with no explicit radial dependence. The superpotential $\tilde{W} (\phi)$ can be expressed as
\be
\tilde{W} (\phi) = \sum_{n \ge 2} w_{n} \phi^n \qquad w_2 = \frac{1}{8} (3 - \Delta_{+})
\ee
and correspondingly 
\be
Y(\phi) =  1 + \sum_{n \ge 2} y_{n} \phi^n
\ee
with 
\be
y_2 = \frac{(3 - \Delta_{+})}{8 (5 - 2 \Delta_{+})}; \qquad
y_3 = \frac{1+ 24 y_2}{(8 - 3 \Delta_{+})} w_3, 
\ee
and so on. Here the cubic counterterm is required for $\Delta_{+} > 8/3$, and there is a corresponding 
logarithmic divergence at $\Delta_{+} = 8/3$ which is cubic in the scalar field. 

For a free scalar in the bulk $w_n = 0$ for $n > 2$, but the expansion of $Y(\phi)$ does not terminate at $n=2$:
\be
Y(\phi) = e^{\frac{1}{4} \phi^2} \sum_{m \ge 0} \frac{(-1)^m}{4^m m!} \frac{\phi^{2m}}{(2 m \Delta_+ - 2 m + 1)}. \label{yphi} 
\ee
However, one should implicitly only retain terms from this series which contribute to divergences. The order $m$ term is required for 
\be
\Delta_{+}  >  3 - \frac{1}{2m}.
\ee
The associated divergence becomes logarithmic at $\Delta_{+} = 3 - 1/2m$: the coefficient at order $m$ in (\ref{yphi}) becomes ill-defined, corresponding to the breakdown of the assumed form of the counterterms. Note that logarithmic terms appear in the asymptotic expansion of the scalar field $\phi$ for half integer conformal dimensions but these are not related to conformal anomalies in the entanglement entropy. 
 
In the $m = 1$ case the regulated onshell action has a logarithmic divergence when $\Delta_{+} = 5/2$
\begin{equation}
  S_{\rm reg} =\frac{1}{4 G_4} \int \dd x \, \sqrt{\tilde{\gamma}} \left( 1 - \frac{1}{4} A_{3 - \Delta_{+}} \varepsilon^{1/2} \log \varepsilon \right) + \ldots
\end{equation}
Here
$A_{1/2} = - \frac{1}{8} \phi_{(0)}^2$ and $\phi = \phi_{(0)} \varepsilon^{1/4} + \ldots$, so we can write this divergence as
\begin{equation}
  S_{\rm reg} = \frac{1}{4 G_4}  \int \dd x \, \sqrt{\tilde{\gamma}} \left( 1 + \frac{1}{32}\phi^2 \log \varepsilon \right) + \ldots.
\end{equation}
The corresponding logarithmic counterterm is then simply
\begin{equation}
  S_{ct} = -\frac{1}{4 G_4} \int \dd x \, \sqrt{\tilde{\gamma}}\frac{1}{32}\phi^2 \log \varepsilon.
\end{equation}
This result is consistent with that of~\cite{Jones2015} who found a logarithmic divergence, in their notation, given by
\begin{equation}
  \delta S = \frac{\mathcal{A}}{8 G_N}(d-2) \lambda^2 h_0 \log(\varepsilon/\varepsilon_{IR})
\end{equation}
which matches our expression under the substitutions $4G_N = 1$, $\mathcal{A} = \int \dd x \, \sqrt{\tilde{\gamma}}$, $d = 3$, $h_0 = \frac{1}{8}$, $\lambda = \phi$ and the relabelling of the cut-off $\varepsilon \to \varepsilon^{1/2}$. This relabelling of the cut off is necessary as theirs is imposed on a $z = \varepsilon$ surface where $ \rho = z^2$.

\bigskip

Thus, to summarise the results of this section, the required counterterms are
\be
S = - \frac{1}{4 G_4} \int_{\partial \Sigma} \dd x \sqrt{\tilde {\gamma}} \left ( 1  +  \frac{3- \Delta_+}{8 (5 - 2 \Delta_{+})} \phi^2 + \cdots \right ), \label{ct-rg}
\ee
where the ellipses denote terms involving higher powers of the scalar field. The counterterm quadratic in scalar fields is necessary for $\Delta_{+} > 5/2$ and is logarithmic at $\Delta_{+} = 5/2$. More generally, new logarithmic divergences involving $n$ powers of
the scalar field arise at 
\be
\Delta_{+} = 3 - \frac{1}{n}
\ee
and an additional counterterm involving $n$ powers of the scalar field is switched on for $\Delta_{+} > 3 - 1/n$. The counterterms can be expressed compactly in terms
of an analytic function of the scalar field $Y(\phi)$ 
\be
S = -  \frac{1}{4 G_4} \int_{\partial \Sigma} \dd x \sqrt{\tilde {\gamma}} Y(\phi), \label{ct-rg2}
\ee
where $Y(\phi)$ is defined in terms of the superpotential for the flow by  \eqref{counterp}. We should emphasise that both expressions \eqref{ct-rg} and \eqref{ct-rg2} are applicable to entangling surfaces in holographic RG flows with flat slicings. For entangling surfaces in generic Einstein-scalar backgrounds there could be additional counterterms dependent on gradients of the scalar field.

\subsection{Entanglement entropy change under relevant perturbation}

In this section we will calculate the change in the renormalized entanglement entropy of a disk entangling region under a small relevant perturbation of the CFT, i.e. we work perturbatively in $\phi_{(0)}$, the source of the relevant operator. As in \cite{Chang:2013mca,Karch2014} it is convenient to express 
the change in the bare entanglement entropy as 
\be
\delta S = \frac{1}{8 G_4} \int \dd^2 x \sqrt{\gamma} T^{mn}_{\rm min} \delta g_{mn}
\ee
where $\gamma$ is the metric on the unperturbed minimal surface, $T^{mn}_{\rm min}$ is the energy momentum tensor for the minimal surface 
\be
T^{mn}_{\rm min} = \gamma^{\alpha \beta} \partial_{\alpha} X^{m}  \partial_{\beta} X^n
\ee
and $\delta g_{mn}$ is the change in the (Einstein) metric induced by the relevant deformation. The latter can always be parameterised as 
\be
ds^2 = \frac{d \rho^2}{4 \rho^2} (1 + \delta f(\rho)) + \frac{1}{\rho} (1 + \delta h (\rho)) dx^{\mu} dx^{\mu},
\ee
and we can furthermore use the gauge freedom to fix $\delta f(\rho) = 0$. The latter gauge choice was implicit in our earlier parameterisation of domain wall geometries.

One can then show that the change in the regulated (bare) entanglement entropy for a disk is
\be
\delta S_{\rm reg} = \frac{\pi R}{4 G_4} \int_{\epsilon}^{R^2} \frac{d \rho}{\rho^{\frac{3}{2}}} (1 + \frac{\rho}{R^2}) \delta h(\rho).
\ee
Note that this expression holds for any small perturbation of the metric which preserves Poincar\'{e} invariance of the dual field theory. 

Working perturbatively in the scalar field amplitude, and taking into account Poincar\'{e} invariance, the most general solution possible for the scalar field
is 
\be
\phi = \phi_{(0)} \rho^{\frac{1}{2}(3 - \Delta_+)} + \phi_{(\Delta_+)} \rho^{\frac{1}{2} \Delta_+}
\ee
(where we assume that $\Delta_{+} \neq 3/2$) and $\phi_{(\Delta_{+})}$  is the normalizable mode of the scalar field. Correspondingly the warp factor is given by 
\be
\delta h = - \frac{1}{8} \left ( \phi_{(0)}^2 \rho^{(3 - \Delta_+)} + \frac{8 \Delta_+}{9} (3 - \Delta_+) \phi_{(0)} \phi_{(\Delta_+)} \rho^{\frac{3}{2}} + \phi_{(\Delta_+)}^2 \rho^{\Delta_+} \right )
\ee
Since we are working perturbatively in the scalar field, we need only retain counterterms which are quadratic in the scalars. In the case of a single scalar field this implies that the only contributing counterterms are those given in (\ref{ct-rg}). 
At $\Delta_+ = 5/2$, the change in the entanglement entropy involves a logarithmic divergence, and thus the renormalized entanglement entropy will be renormalization scheme dependent. 

The change in the renormalized entanglement entropy is hence (for $\Delta_{+} \neq 5/2$)
\bea
\delta S_{\rm ren} &=&  \frac {\pi }{16 (2 \Delta_+ - 5) G_4} \phi_{(0)}^2 R^{2 (3- \Delta_+)}  \label{shiftR} \\
&& + \frac{\pi}{36 G_4} \Delta_+ (\Delta_+  -3 )  \phi_{(0)} \phi_{(\Delta_+)} R^3 + \frac{\pi}{16 (2 \Delta_+-1) G_4} \phi_{(\Delta_+)}^2 R^{2 \Delta_+}. \nn
\eea
Working to quadratic order in the scalar field one cannot impose regularity in the bulk as $\rho \rightarrow \infty$ as both modes are unbounded. 
On dimensional grounds, however, 
\be
\phi_{(\Delta)} \propto \phi_{(0)}^{\frac{\Delta_+}{(3 - \Delta_+)}}
\ee
for $\frac{3}{2} < \Delta_+ < 3$. Hence $\phi_{ (\Delta_+ )} \sim \phi_{(0)}^{\delta}$ with $\delta > 1$, and the normalizable mode is subleading in powers of the non-normalizable mode, as we will see in the full solution given in the next section.

Therefore
\be
\delta S_{\rm ren} =  \frac {\pi }{16 (2 \Delta_+ - 5) G_4} \phi_{(0)}^2 R^{2 (3- \Delta_+)} + \cdots \label{rgflow}
\ee
where ellipses denote terms which are of higher order in the source. This quantity is positive for $\Delta_+ > 5/2$ but negative for relevant deformations with $3/2 < \Delta_+ < 5/2$. Recalling that the F quantity is proportional to minus the renormalized entanglement entropy the change in the F quantity is {\it positive} for relevant deformations with $3/2 < \Delta_{+} < 5/2$. A related result was obtained in \cite{Nishioka2014}, although the sign of the quantity was not explicitly identified in that work. 

For operators of dimension $\Delta_- < 3/2$, the non-normalizable mode $\phi_{(0)}$ is not the operator source: the correct source is obtained from a Legendre transformation of the onshell action \cite{Klebanov:1999tb}. Such a Legendre transformation cannot be carried out without working to higher orders in the non-normalizable mode  $\phi_{(0)}$ and thus we cannot obtain the entanglement entropy for this case without knowledge of the higher order solution.

Now let us consider the special case of $\Delta = 3$, i.e. marginal operators. In this case the warp factor is unchanged by the non-normalizable mode of the scalar field $\phi_{(0)}$, i.e. integrating the equations of motion we obtain
\be
\delta h = - \frac{1}{8} \phi_{(3)}^3 \rho^3.
\ee
Since the non-normalizable mode does not affect the metric, there are no new divergences and no counterterms depending on the scalar field. At $\Delta = 3$ there are also no
possible finite counterterms since the finite counterterm
\be
S_{\rm ct} = - \frac{1}{4 G_4} \int dx \sqrt{\gamma^h} \left ( {\cal K} \phi^2 \right )
\ee
does not respect the complementarity requirement.
The renormalized entanglement entropy for a marginal deformation is thus
\be
\delta S_{\rm ren} =  
 \frac{\pi}{80G_4} \phi_{(3)}^2 R^{6}. \nn
\ee
In the vacuum of the marginally deformed conformal field theory $\phi_{(3)} = 0$ and the change in the renormalised entanglement entropy is therefore zero for the vacuum of the deformed conformal field theory. Note that the change in the renormalized entanglement entropy is hence implicitly not analytic in the operator dimension as 
$\Delta \rightarrow  3 $; this is however permissible, since the spectrum of operators is discrete. 

\bigskip

We can also compute the change in the quantity $F(R)$ defined in section \ref{two}:
\be
\delta F (R) = - \delta S_{\rm reg}(R) + R \frac{\partial S_{\rm reg}(R)}{\partial R}
\ee
For $\Delta < 3$ 
\be
\delta F(R) = - \frac{\pi \phi_{(0)}^2}{16 G_4} R^{6 - 2 \Delta}
\ee
which is negative for all relevant deformations. This does not agree numerically with $\delta S_{\rm ren}$, but it is the latter which is by construction related to the renormalized F quantity by the CHM map. 

For $\Delta = 3$, the change in the regulated entanglement entropy is zero, as the metric is unchanged, and therefore 
\be
\delta F(R) = 0.
\ee
Note that implicitly the change in $\delta F(R)$ is therefore also non-analytic at $\Delta \rightarrow 3$. 

\bigskip

It may seem surprising that the F quantity decreases along RG flows generated by operators of dimensions $3/2 < \Delta_{+} < 5/2$. 
The results discussed above actually follow directly from the subadditivity property of the (regularised) entanglement entropy: recall that the latter implies that $\partial^2 S_{\rm reg}/\partial R^2 \le 0$. Our analysis implies that the counterterms scale with the size of the entangling region, i.e. $S_{\rm ct} \propto R$. Therefore subadditivity implies 
\be
\frac{\partial^2 S_{\rm ren}}{\partial R^2} = \frac{\partial^2 S_{\rm reg}}{\partial R^2} \le 0.
\ee
However, on dimensional grounds, when we work to quadratic order in the source $S_{\rm ren}$ must take the form
\be
S_{\rm ren} = - \frac{\pi}{2 G_4} + a_{2(3 - \Delta_{+})} \phi_{(0)}^2 R^{2 (3 -\Delta_{+})} + \cdots 
\ee
for $\Delta_{+} > 3/2$ where $a_{2(3 - \Delta_{+})}$ is a dimensionless constant. Here we use the explicit form for the 
leading term, which is independent of $R$.  Differentiating twice with respect to $R$ then gives
\be
\frac{\partial^2 S_{\rm ren}}{\partial R^2} = 2 (3 - \Delta_{+})(5 - 2 \Delta_{+}) a_{2(3 - \Delta_{+})}  \phi_{(0)}^2 R^{2 (2 - \Delta_{+})} + \cdots
\ee
This is negative semi-definite (as required by strong subadditivity) provided that 
\be
(3 - \Delta_{+})(5 - 2 \Delta_{+}) a_{2(3 - \Delta_{+})} \le 0,
\ee
i.e. provided that $a_{2 (3 - \Delta_{+})} \ge 0$ for $\Delta_{+} \ge 5/2$ and  $a_{2 (3 - \Delta_{+})} \le 0$ for $\Delta_{+} \le 5/2$, as we found above. 

\bigskip

A related result is found by
directly computing the change in the free energy to quadratic order in the source for holographic RG flows on a sphere driven by the same operators. Deformations of the theory on the sphere
\be
I_{\rm CFT} \rightarrow I_{\rm CFT} + \int_{S^3} d^3 \Omega \; \psi_{(0)} {\cal O}_{\Delta_+}, \label{def1}
\ee
where the source $\psi_{(0)}$ is independent of the spherical coordinates and $d\Omega$ is the measure on the $S^3$,
may be described holographically by spherical sliced domain walls. Again working to quadratic order in the source, the change in the free energy is positive for operators of dimensions $3/2 < \Delta_{+} < 5/2$ \cite{Taylor1}. 

Note that such deformations are not equivalent to conformal transformations of the holographic RG flows considered here, which are dual to deformations of the theory on flat space:
\be
I_{\rm CFT} \rightarrow I_{\rm CFT} + \int_{R^3} d^3x \phi_{(0)} {\cal O}_{\Delta_{+}}. \label{def2}
\ee
To understand this point further, it is useful to recall the relationship between spherical and Poincar\'{e} coordinates for anti-de Sitter. The former can be described in terms of the following embedding into $R^{1,4}$:
\be
X^0 = \cosh w \qquad X^1 + i X^2 = \sinh w \cos \theta e^{i \tau_E} \qquad X^3 + i X^4 = \sinh w \sin \theta e^{i \phi}
\ee
so that 
\be
ds^2 = dw^2 + \sinh^2 w \left ( d \theta^2 + \cos^2 \theta d \tau_E^2 + \sin^2 \theta d \phi^2 \right ).
\ee
Poincar\'{e} coordinates can be obtained by setting
\bea
X^0 + X^1 &=& \frac{1}{\rho^{1/2}} \qquad
 X^0 - X^1 = \left ( \rho^{1/2} + \frac{1}{\rho^{1/2}} (t_E^2 + x^2 + y^2) \right ) \\
 X^2 &=& \frac{t}{\rho^{1/2}} \qquad
 X^3 = \frac{x}{\rho^{1/2}} \qquad
 X^4 = \frac{y}{\rho^{1/2}}, \nn
\eea
resulting in 
\be
ds^2 = \frac{d\rho^2}{4 \rho^2} + \frac{1}{\rho} \left (dt_E^2 + dx^2 + dy^2 \right ). 
\ee
From these relations it is clear that the radial coordinate in spherical slicings, $w$, depends on both $\rho$ and $|x| \equiv (t_E^2 + x^2 + y^2)^{\frac{1}{2}}$. Conversely
the Poincar\'{e} radial coordinate $\rho$ depends on $(w,\theta,\tau_{E})$. Therefore flows which depend only on $w$ or $\rho$, respectively, are not equivalent to each 
other: a flow which depends only on $w$ will depend on the Poincar\'{e} norm $|x|$ as well as $\rho$. 

From the field theory perspective, the theories on the $S^3$ and on $R^3$ are related by the conformal transformation described earlier, with the relevant conformal factor being 
given by (\ref{chm5}). While the original conformal field theory is of course unaffected by this conformal factor, mapping (\ref{def2}) to the sphere results in 
\be
\phi_{(0)} \rightarrow  \Omega^{\Delta_{+} - 3} (\theta, \tau_E) \phi_{(0)},
\ee
i.e. the transformed source is not homogeneous over the $S^3$, and therefore the deformations on $S^3$ and $R^3$ by homogeneous sources are not conformally equivalent. Thus,
while the change in the renormalized entanglement entropy is indeed related to a change in the free energy on the $S^3$, the latter is the change under a deformation which breaks the $SO(4)$ invariance.

\subsection{Top down RG flow}

Let us now consider entanglement entropy in holographic RG flows which have top down embeddings. We will discuss the following single scalar example, taken from \cite{Papadimitriou:2006dr}. Let the potential be 
\be
V(\phi) = 6 \cosh \left (\frac{\phi}{\sqrt{3}} \right )
\ee
which arises in a consistent truncation of ${\cal N} = 8$ gauged supergravity, which in turn is a consistent truncation of M theory compactified on $S^7$. 
The RG flow equations can be used to construct analytic domain wall solutions in which the metric is conveniently expressed as 
\be
ds^2 = \frac{ (1 + \nu r + \sqrt{1 + 2 \nu r + r^2})}{2 r^2 \sqrt{1- r^2} (1 + 2 \nu r + r^2)} d r^2 + 
\frac{\sqrt{1-r^2}}{2 r^2} (1 + \nu r + \sqrt{1 + 2 \nu r + r^2}) dx^{\mu} dx_{\mu}
\ee
and the scalar field profile is 
\be
\phi = \sqrt{3} \tanh^{-1} (r).
\ee
The parameter $\nu \ge -1$ is arbitrary with $\nu = -1$ corresponding to a supersymmetric domain wall of the supergravity theory. Here $r \rightarrow 0$ corresponds to the conformal boundary. Note that in all cases the metric has a singularity at $r = 1$; this singularity is null in the supersymmetric case and timelike in all other cases but the singularity is good according to the standard criteria. 
The scalar mass associated with the potential is $M^2 = -2$, which corresponds to the cases of $\Delta_- = 1$ and $\Delta_+ = 2$, i.e. the mass is such that both quantisations are possible and mixed boundary conditions can be considered. 

We can reintroduce the scalar field amplitude as a parameter by letting 
\be
r = c \tilde{r}; \qquad
x^{\mu} = c \tilde{x}^{\mu}
\ee
so that 
\bea
ds^2 &=& \frac{ (1 + \nu c \tilde{r} + \sqrt{1 + 2 \nu c \tilde{r} + c^2 \tilde{r}^2})}{2 \tilde{r}^2 \sqrt{1- c^2 \tilde{r}^2} (1 + 2 \nu c \tilde{r} + c^2 \tilde{r}^2)} d \tilde{r}^2 + 
\frac{\sqrt{1- c^2 \tilde{r}^2}}{2 \tilde{r}^2} (1 + \nu c \tilde{r} + \sqrt{1 + 2 \nu c \tilde{r} + c^2 \tilde{r}^2}) d\tilde{x}^{\mu} d\tilde{x}_{\mu} \nn \\
\phi &=& \sqrt{3} \tanh^{-1} (c \tilde{r}). 
\eea
We can then change coordinates for $ c \tilde{r} \ll 1$ as
\be
\tilde{r}^2 = \rho + \nu c \rho^{\frac{3}{2}} + \cdots 
\ee
to obtain 
\bea
ds^2 &=& \frac{d \rho^2}{4 \rho^2} + \frac{1}{\rho} (1 - \frac{3}{8} c^2 \rho + \cdots  ) d\tilde{x}^{\mu} d\tilde{x}_{\mu};  \label{fg-exp-rg}  \\
\phi &=& \sqrt{3} c \left ( \rho^{\frac{1}{2}} +\frac{1}{2} \nu c \rho + \cdots \right ) \nn
\eea
from which we can read off that
\be
\phi_{(0)} = \sqrt{3} c; \qquad
\phi_{(\Delta_+)} \equiv \phi_{(1)} = \frac{1}{2} \sqrt{3} \nu c^2,
\ee
i.e. the normalizable mode is of order the non-normalizable mode squared. (This had to be true on dimensional grounds in a solution which depends on only one dimensionful parameter, $c$.)
Thus substituting into \eqref{shiftR} we obtain
\be
\delta S_{\rm ren} = - \frac{\pi}{48 G_4} \phi_{(0)}^2 R^4 + {\cal O} \left ( \phi_{(0)}^3 \right ), \label{smallR}
\ee
in agreement with (\ref{rgflow}) in the case of $\Delta_+ = 2$. 

\bigskip

The result (\ref{smallR}) can be interpreted as follows. There are only two physical scales in the field theory: the source for the operator deformation $c$ and the size of the entangling region $R$. When $c R \ll 1$, the entangling surface is small and does not penetrate far into the bulk. The region probed by the entangling surface is well-described by the asymptotic Fefferman-Graham expansion (\ref{fg-exp-rg}), and therefore one can use the results of the previous section to compute the entanglement entropy. Note that the result does not depend on the parameter $\nu$, i.e. it is same for supersymmetric and non-supersymmetric RG flows. 

Now consider increasing the radius of the entangling surface at fixed source. On dimensional grounds $\delta S_{\rm ren}$ is a function of $\phi_{(0)} R$.  Since 
$\partial^2 S_{\rm ren}/\partial R^2 \le 0$, $\partial \delta S_{\rm ren}/\partial R$ must decrease monotonically with the radius $R$ and $\delta S_{\rm ren}$ must be negative for all $R$. 

\section{Renormalization via the replica trick} \label{five}

In the previous sections we have described a renormalization procedure for entanglement entropy which is based on the holographic realisation of entanglement entropy in terms of minimal 
surfaces. It is difficult to translate this procedure directly into a field theoretic definition of renormalization, since the Ryu-Takayanagi functional itself does not follow directly from field theory. 

A conceptual derivation of the Ryu-Takayanagi functional has been obtained by Lewkowycz-Maldacena \cite{Lewkowycz:2013nqa} via the replica trick. 
The entropy associated with a density matrix $\rho$ is expressed as
\be
S = -n \partial_n \left [ \log Z(n) - n \log Z(1) \right ]_{n=1} \label{rep1}
\ee
where $Z(n) = {\rm Tr} (\rho^n)$ and $Z(1)= {\rm Tr} (\rho)$ is the usual partition function. If we are interested in the entropy of a thermal state, then $Z(n)$ is constructed by extending the period of the thermal circle by a factor of $n$. In the case of entanglement entropy, $Z(n)$ is constructed by extending the period of the circle around the boundary of the entangling region by a factor of $n$, where implicitly $n$ is an integer. Assuming that the resulting expression is analytic in $n$, one can obtain the entropy by analytically continuing to $n=1$. 

Holographically $Z(n)$ can be computed in terms of the Euclidean actions:
\be
S = n \partial_n \left [ I(n) - n I(1) \right ]_{n=1}. \label{rep2}
\ee
Here $I(1)$ represents the onshell Euclidean action for the bulk geometry while $I(n)$ represents the onshell Euclidean action for the replica bulk geometry.
For a thermal state, the bulk geometry associated with $Z(1)$  is a black hole and the replica is constructed by extending the period of the thermal circle by a factor of $n$. 
It was shown by Lewkowycz-Maldacena \cite{Lewkowycz:2013nqa} that for a bulk theory described by Einstein gravity (\ref{rep1}) then localises on the horizon of the black hole, i.e. 
\be
S = \frac{A}{4 G_{d+1}}.
\ee
In particular, the volume divergences of the onshell actions (associated with UV divergences in the field theory) by construction cancel, since the replica geometry asymptotically 
matches $n$ copies of the original geometry. 

For the entanglement entropy, the bulk geometry associated with $Z(1)$ corresponds to the usual bulk dual of the given state in the field theory. 
The replica is constructed by extending the period of the circle around the entangling region boundary by a factor of $n$. Following the same logic as in Lewkowycz-Maldacena,
the expression \eqref{rep2} localises on the minimal surface corresponding to the extension of the boundary of the entangling region into the bulk (see the discussions in \cite{Jones:2016iwx}).  However, unlike the black 
hole case, the volume divergences of the bulk actions in \eqref{rep2} do not cancel, as the entangling surface itself has area divergences. 

We can formally write down a renormalized entanglement entropy as 
\be
S_{\rm ren} = n \partial_n \left [ I_{\rm ren}(n) - n I_{\rm ren}(1) \right ]_{n=1} \label{rep3}
\ee
where the quantities appearing on the right hand side are the renormalized bulk actions. Equivalently, 
\be
S_{\rm ct} = n \partial_n \left [ I_{\rm ct}(n) - n I_{\rm ct}(1) \right ]_{n=1} \label{rep4}
\ee
Let us first focus on the specific case of entangling surfaces in $AdS_4$, for which the usual counterterms for the onshell action are \cite{DeHaro2001}
\be
I_{\rm ct}(1) = \frac{1}{4 \pi G_4} \int_{\partial {\cal M}} d^3 x \sqrt{h} \left (- \frac{1}{2} K + 1 + \frac{1}{4} R \right ). 
\ee
Here we define the bulk geometry to be ${\cal M}$ and its boundary to be $\partial {\cal M}$, and $K$ denotes the trace of the extrinsic curvature of $\partial {\cal M}$ embedded into 
${\cal M}$. (The first term is the usual Gibbons-Hawking term.)

Since the replica geometry is also asymptotically locally $AdS_4$, the counterterms are
\be
I_{\rm ct}(n) = \frac{1}{4 \pi G_4} \int_{\partial {\cal M}_n} d^3 x \sqrt{h_n} \left (- \frac{1}{2} K_n  + 1 + \frac{1}{4} R_n \right ).
\ee
where $h_n$ is the boundary metric for the replica geometry and $K_n$ and $R_n$ are the associated extrinsic curvature and Ricci scalar, respectively. 
Now the replica geometry by construction matches the original 
geometry except at the fixed point set of $\partial_{\tau}$, where $\tau$ is the circle around the boundary of the entangling region and its extension into the bulk. 
At this fixed point set the metric and the extrinsic curvature of the replica match the original metric, but the intrinsic curvature invariants of the replica
receive contributions from the conical singularity. In the case of interest $R= 0$ but in the replica geometry due to the conical singularity
\be
\int d^3 x \sqrt{h_n} R_n  = 4 \pi  (1-n) \int_{\partial \Sigma} dx \sqrt{\tilde{\gamma}}
\ee
and hence we find that 
\be
S_{\rm ct} = -\frac{1}{4 G_4} \int_{\partial \Sigma} dx \sqrt{\tilde{\gamma}},
\ee
which matches the counterterm obtained by our explicit calculations in section \ref{three}. 

For the case of entangling surfaces in holographic RG flows the counterterms to quadratic order in the scalar field are  \cite{DeHaro2001}
\be
I_{\rm ct}(1) = \frac{1}{4 \pi G_4} \int_{\partial {\cal M}} d^3 x \sqrt{h} \left (1 + \frac{1}{16} (3 - \Delta_+) \phi^2 + \frac{1}{4} R + \frac{\Delta_{+} - 3}{32 (2 \Delta_{+} - 5)} R \phi^2 \right ),
\ee
where we drop the Gibbons-Hawking term as it does not contribute to the entanglement entropy counterterms, and we 
 also neglect terms involving derivatives of the scalar field, i.e. we restrict to homogeneous scalar field configurations. 
Following the same steps as above, we can then show that
\be
S_{\rm ct} = - \frac{1}{4 G_4} \int_{\partial \Sigma} dx \sqrt{\tilde{\gamma}} \left ( 1 + \frac{\Delta_{+} - 3}{8 (2 \Delta_{+} - 5)} \phi^2 \right )
\ee
which is again in agreement with our explicit results of section \ref{four}. 

\bigskip

Let us now move to general dimensions. For an asymptotically locally $AdS_{D+2}$ spacetime the counterterms are \cite{DeHaro2001}
\bea
I_{\rm ct}(1) &=& \frac{1}{16 \pi G_{D+2}} \int_{\partial {\cal M}} d^{D+1} x \sqrt{h} \left (2 D + \frac{1}{(D-1)} R   \right . \\
&& \qquad  \qquad \left . + \frac{1}{(D-3)(D-1)^2} \left (R_{ab} R^{ab} - \frac{D+1}{4 D} R^2 \right )  + \cdots  \right ). \nn
\eea
This expression should be understood as containing only the appropriate divergent terms in any given dimension; moreover, for odd $D$ there are logarithmic 
counterterms. In particular, for $D=3$  the third counterterm is replaced by the logarithmic counterterm 
\be
 \frac{1}{16 \pi G_{5}} \int_{\partial {\cal M}_{\varepsilon}} d^{4} x \sqrt{h} \frac{1}{8} \left (R_{ab} R^{ab} - \frac{1}{3} R^2 \right ) \ln \varepsilon.
\ee

In the replica geometry, the contributions to the curvature from the conical singularity are given by \cite{Solodukhin:2008dh}
\bea
R_{n} &=& R + 4 \pi (1 - n ) \delta_{\partial \Sigma} + {\cal O}(1-n)^2; \label{curve1} \\
R_{n ab} &=& R_{ab} + 2 \pi (1 - n) n_a n_b \delta_{\partial \Sigma} + {\cal O}(1-n)^2, \nn
\eea
where $\delta_{\partial \Sigma}$ is a delta function localised on the entangling surface. Here $n_a^k$ with $k=1,2$ represent orthonormal vectors to the entangling surface and
\be
n_a n_b = \sum_k n^k_a n^k_b.
\ee
Following the same steps as above, we can immediately read off the leading counterterm for the entanglement entropy as 
\be
S_{\rm ct,1} = -  \frac{1}{4 (D-1) G_{D+2}} \int_{\partial \Sigma} d^{D-1} x \sqrt{\tilde{\gamma}}, 
\ee
in agreement with our earlier result. 

For the higher order counterterms, one can use the following expressions \cite{Fursaev:2013fta}
\bea
\int_{\partial {\cal M}_n} d^{D+1} x \sqrt{h_n} R_{n}^2 &=& n \int_{\partial {\cal M}} d^{D+1} x \sqrt{h} R^2 + 8 \pi (1-n)  
\int_{\partial \Sigma} d^{D-1} x  \sqrt{\tilde{\gamma}} R  \label{fur1}  \\
\int_{\partial {\cal M}_n} d^{D+1} x \sqrt{h_n} R_{nab} R_n^{ab} &=& n \int_{\partial {\cal M}}  d^{D+1} x \sqrt{h} R_{ab} R^{ab} + 4 \pi (1-n)  
\int_{\partial \Sigma} d^{D-1} x  \sqrt{\tilde{\gamma}} (R_{ii} - \frac{1}{2} k^2), \nn
\eea
where implicitly we work to leading order in $(1-n)$ and we define 
\be
k^2 = \sum_k ({\cal K}^{k})^2
\ee
with $R_{ii}$ corresponding to invariant projections of the Ricci tensor onto the subspace orthogonal to $\partial \Sigma$, see \cite{Solodukhin:2008dh}. 

In section \ref{three}, we analysed the entanglement entropy counterterms assuming that the entangling surface is static and that the curvature of the boundary metric is zero.
In such a case $R_{ii} = R = 0$ and the extrinsic curvature in the time direction is zero. Thus the second counterterm becomes
\be
S_{\rm ct,2} =  - \frac{1}{8 (D-1)^2 (D-3) G_{D+2}} \int_{\partial \Sigma} d^{D-1} x \sqrt{\tilde{\gamma}}{\cal K}^2,
\ee
where ${\cal K}$ refers to the trace of the extrinsic curvature of the surface embedded into a constant time hypersurface. Similarly in $D=3$ the logarithmic counterterm is
\be
S_{\rm ct,2} =  \frac{1}{64 G_{5}} \int_{\partial \Sigma} d^{3} x \sqrt{\tilde{\gamma}} {\cal K}^2 \ln \varepsilon,
\ee
which is in agreement with the expression obtained in \cite{Solodukhin:2008dh} for the anomaly in the entanglement entropy for 4d CFTs with a holographic dual. (See \cite{Solodukhin:2008dh} for the conformal
anomaly in a general 4d conformal field theory in which $a \neq c$.)

One can now immediately generalize the entanglement entropy counterterms to the case of a general embedding into a curved boundary metric obtaining
\bea
S_{\rm ct} &=& -  \frac{1}{4 (D-1) G_{D+2}} \int_{\partial \Sigma} d^{D-1} x \sqrt{\tilde{\gamma}} \\
&& \qquad -  \frac{1}{4 (D-1)^2(D-3)  G_{D+2}} \int_{\partial \Sigma} d^{D-1} x \sqrt{\tilde{\gamma}} \left ( R_{ii} - \frac{1}{2} k^2 - \frac{D+1}{2D} R \right ), \nn
\eea
where one can use the Gauss-Codazzi relations to write $R_{ii}$ and $R$ in terms of intrinsic and extrinsic curvatures of $\partial \Sigma$. 

\subsection{Higher derivative generalizations}

Using the replica trick, we can derive the renormalized entanglement entropy functional from any higher derivative gravity for which the renormalized bulk action is known. Let
us consider the particular example of Gauss-Bonnet gravity, with bulk action 
\be
I = - \frac{1}{16 \pi G_{D+2}} \int_{{\cal M}} d^{D+2} x \sqrt{g} \left [ R_g + D (D+1) + \lambda \left (R_{mnpq} R^{mnpq} - 4 R_{mn} R^{mn} + R_g^2 \right ) \right ]
\ee
where $\lambda$ is the Gauss-Bonnet coupling. 

One can derive the entanglement entropy functional by the replica trick used above, see \cite{Fursaev:2013fta,Bhattacharyya:2013jma}, using the bulk versions of (\ref{fur1}) together with the additional relation 
\bea
\int_{{\cal M}_n} d^{D+2} x \sqrt{g} R_{mnpq} R^{mnpq} &=& n \int_{{\cal M}_n} d^{D+2} x \sqrt{g} R_{mnpq} R^{mnpq} \\
&& \qquad + 8 \pi (1-n) \int_{\Sigma} d^{D} y \sqrt{\gamma} \left (R_{ijij} - {\rm Tr}(k^2) \right ), \nn
\eea
where we neglect terms of higher order in $(n-1)$ and $R_{ijij}$ denotes the projection of the Riemann tensor in the directions orthogonal to the entangling surface. Also
\be
{\rm Tr}(k^2) = \sum_{k=1}^{2} {\cal K}^k_{ab} {\cal  K}^{k  ab}.
\ee
Thus the entanglement entropy functional consists of the usual Ryu-Takayanagi term plus additional terms 
\bea
S = \frac{1}{4 G_{D+2}} \int_{\Sigma} d^D y \sqrt{\gamma} + \frac{\lambda}{G_{D+2}} \int_{\Sigma} d^D y \sqrt{\gamma} \left (  R_{ijij} -  {\rm Tr}(k^2) - 2 R_{ii} + k^2 + R \right  ),
\eea
where implicitly all terms can be written in terms of extrinsic and intrinsic curvatures on the entangling surface.
As shown in \cite{Fursaev:2013fta}, in five bulk dimensions the latter term can be simplified using the Gauss-Codazzi relations to give
\be
S = \frac{1}{4 G_{5}} \int_{\Sigma} d^3 y \sqrt{\gamma} \left  (1 + 2 \lambda \hat{R} \right ),
\ee
with $\hat{R}$ the intrinsic curvature of the entangling surface. 

Now the bulk equations of motion admit as $AdS_5$ as a solution, but the radius of the $AdS_5$ depends on the Gauss-Bonnet coupling, i.e. the $AdS_5$ metric is
\be
ds^2 = l^2 (\lambda) \left ( \frac{d \rho^2}{4 \rho^2} + \frac{1}{\rho} dx \cdot dx \right )
\ee
where the radius is given by
\be
l^4 (\lambda) - l^2(\lambda) + 2 \lambda = 0.
\ee
One can then straightforwardly show that the leading order counterterm for the entanglement entropy is given by 
\be
S_{\rm ct} = -  \frac{1}{8 G_{5}} \int_{\partial \Sigma} d^{2} x \sqrt{\tilde{\gamma}} \left ( l(\lambda) - 12 \frac{\lambda}{l(\lambda)} \right ),
\ee
where we use the fact that the entangling surface is asymptotically locally hyperbolic and thus $\hat{R} = - 6//l(\lambda)^2 + \cdots$. There is also a subleading logarithmic divergence associated with the conformal anomaly; this is known from the work of \cite{Solodukhin:2008dh}. 

Now the leading order counterterm for the entanglement entropy is inherited from the subleading counterterm for the bulk action, i.e. the counterterm 
\be
I_{ct} = a_2 \int d^4 x \sqrt{h} R.
\ee
This counterterm is not known to all orders in $\lambda$, although it was derived perturbatively in $\lambda$ in \cite{Liu:2008zf,Cremonini:2009ih,Jahnke:2014vwa}. The relation with entanglement entropy immediately gives the coefficient of this counterterm to be 
\be
a_2 = \frac{1}{32 G_5}  \left ( l(\lambda) - 12 \frac{\lambda}{l(\lambda)} \right ),
\ee
i.e. the entanglement entropy counterterms provide a quick method of deriving or checking counterterms in the bulk action involving the curvature.

\subsection{Domain walls}

In this section we show how the counterterms for asymptotically locally $AdS$ solutions of a theory with a single scalar imply the entanglement entropy counterterms discussed
in section \ref{four}. The bulk Euclidean action is 
\be
I = - \frac{1}{16 \pi G_{4}} \int_{{\cal M}} d^{4} x \sqrt{g} \left ( R_g - \frac{1}{2} (\partial \phi)^2 + V(\phi) \right ).
\ee
In general the counterterms for asymptotically locally $AdS$ solutions of this action can be expressed in the form
\be
I_{ct} = \frac{1}{16 \pi G_4} \int_{\partial {\cal M}} d^3 x \sqrt{h} \left ( {\cal W} (\phi)  + {\cal Y} (\phi) R  + \cdots \right ),
\ee
where ${\cal W}(\phi)$ and ${\cal Y}(\phi)$ are analytic functions of the scalar field $\phi$. Here the ellipses denote terms which depend on gradients of the scalar field; as in the discussions above, such terms are not relevant when using the replica trick to derive the entanglement entropy counterterms. In the above expression we assume generic values of the dual operator dimension such that there are no conformal anomalies; for specific values of the operator dimension there will however be conformal anomalies. 

For a flat domain wall solution, characterized by a given superpotential $W(\phi)$, the only contributing counterterm is ${\cal W}(\phi) = 4 W (\phi)$, since in this case $R = 0$. 
To use the replica trick we need to know how the counterterms for the bulk action depend on the curvature of the boundary metric i.e. we cannot restrict to flat sliced domain walls: the counterterm for the entanglement entropy follows from the term involving the Ricci scalar above, i.e.
\be
S_{\rm ct} = - \frac{1}{4 G_4} \int_{\partial \Sigma} dx \sqrt{\tilde{\gamma}} {\cal Y}(\phi).
\ee

We can understand the specific form of ${\cal Y}(\phi)$ for entanglement entropy in a flat sliced domain wall as follows. 
We begin with solutions of the equation of motion correspondins to domain walls with homogeneous slicing, i.e. 
the metric is
\be
ds^2 = dw^2 + e^{2 {\cal A}(w)} d \Omega_3^2
\ee
and the scalar field profile is $\phi(w)$. We let the Ricci scalar of the slicing be $\hat{r}$ where, for example, $\hat{r} = 6$ for unit radius spherical slices. The equations of motion are then
\bea
\ddot{\phi} + 3 \dot{\phi} \dot{\cal A} &=& - V'(\phi); \\
- \frac{\hat{r}}{6 } e^{-2 {\cal A}} - \frac{1}{4} \dot{\phi}^2 &=& \ddot{\cal A}. \nn
\eea
These equations are identical to those discussed in section \ref{four}, apart from the curvature contribution to the second equation. 

Now let us work in the limit that $\hat{r} \ll 1$. For $\hat{r}  = 0$, the equations admit the first order form discussed in section \ref{four}, in terms of the superpotential $W(\phi)$. For $\hat{r} \ll 1$, the equations of motion are solved to order $\hat{r}^2$ by
\be
\dot{\cal A} = W \qquad
\dot{\phi} = - 4 \frac{d W}{d \phi} + \hat{r} f(\phi)
\ee
provided that 
\bea
3 W f(\phi) - 4 \frac{d}{d \phi} \left ( f(\phi) \frac{d W}{d \phi} \right )  &=& 0; \\
f(\phi)\frac{d W}{d \phi} = \frac{1}{6}  e^{-2 {\cal A}}. \nn
\eea
The regulated onshell action (including the Gibbons-Hawking term) thus becomes
\bea
I_{\rm reg} &=& - \int^R dw \int d \Omega_3 \left ( e^{\cal A} \hat{r} + {\cal O}(\hat{r}^2) \right ) - \frac{1}{4 \pi G_4} \int d \Omega_3 \left [ e^{3 \cal A} W  \right ]_R, \\
&=& - \frac{1}{16 \pi G_4} \int^R dw \int d \Omega_3 \left ( e^{\cal A} \hat{r} + {\cal O}(\hat{r}^2) \right ) - \frac{1}{4 \pi G_4} \int_{\partial M} d^3 x \sqrt{h} W, \nn
 \eea
where we have used the field equations to linear order in $\hat{r}$ and in the second line we write the boundary term in covariant form. The bulk term can be expressed
as a covariant boundary term
\be
- \frac{1}{16 \pi G_4} \int_{\partial M} d^3 x \sqrt{h} R Y(\phi)
\ee
provided that 
\be
\frac{d}{dw} \left ( \sqrt{h} R Y(\phi) \right ) = e^{\cal A} \hat{r}.
\ee
However, 
\be
\frac{d}{dw} \left ( \sqrt{h} R Y \right ) = \hat{r} \frac{d}{dw} \left ( e^{\cal A} Y \right ) =  e^{\cal A} \hat{r} \left (W Y- 4 \frac{d W}{d \phi} \frac{d Y}{d \phi} \right ),
\ee
where we drop terms of higher order in $\hat{r}$ and use the field equations. Therefore the required counterterms are 
\be
I_{\rm ct} = \frac{1}{16 \pi G_4} \int_{\partial M} d^3 x \sqrt{h} \left ( 4 W + R Y \right ), \label{ct4}
\ee
with 
\be
\left (W Y- 4 \frac{d W}{d \phi} \frac{d Y}{d \phi} \right ) = 1,
\ee
as we found in section \ref{four}. Note that terms of higher order in $\hat{r}$ would not contribute to the counterterms, as they do not give rise to divergent terms. 

We calculated the curvature term in \eqref{ct4} by working with a homogeneous domain wall. To use the replica trick, we need to consider a replica space in which
the curvature of the boundary is given by \eqref{curve1}, in the limit that $n \rightarrow 1$, i.e. it is not homogeneous, but \eqref{ct4} is covariant and still applies. 
Note that the slices of the domain wall are flat, up to conical singularity terms which are proportional to $(n-1)$, and hence $R$ is small as $n \rightarrow 1$. It is therefore
indeed true to leading order in $(n-1)$ that the replica geometry is still governed by the superpotential $W$. Following the same steps as earlier in this section, we can then
immediately read off the counterterm action for the entanglement entropy as 
\be
S_{\rm ct} = -  \frac{1}{4 G_4} \int_{\partial \Sigma} dx \sqrt{\tilde{\gamma}} Y(\phi),
\ee
as we found in section \ref{four}. 

It is important to note that this expression holds specifically for flat domain wall geometries associated with a superpotential $W$. A generic curved domain wall geometry is not governed by a single real superpotential (see \cite{Skenderis:2006jq,Papadimitriou:2011qb}) and the analysis above would need to be generalized for such cases. 

\section{Conclusions} \label{six}

In this paper we have shown how the holographic entanglement entropy may be renormalized using appropriately covariant boundary counterterms. This renormalization procedure is inherited directly from the renormalization of the partition function, using the replica trick. 

We analysed renormalization for entangling surfaces in asymptotically locally AdS spacetimes in any dimension and in flat sliced holographic RG flows in four bulk dimensions. We also showed that the renormalization procedure can be extended to higher derivative theories such as Gauss-Bonnet. It would be straightforward to generalize our results to include entangling surfaces with cusps and to non-conformal holographic setups using \cite{Kanitscheider:2008kd}. It would be interesting to explore real-time holography in the context of entanglement entropy, using the techniques of \cite{Skenderis2009} for the HRT functional \cite{Hubeny:2007xt}. 
 
While it is difficult to relate the area renormalization of the holographic entanglement entropy functional directly to field theory renormalization, the replica trick expresses our renormalised entanglement entropy in terms of renormalized partition functions, i.e.
\be
S_{\rm ren} = -n \partial_n \left [ \log Z_{\rm ren}(n) - n \log Z_{\rm ren}(1) \right ]_{n=1}. \label{rep6a}
\ee
This expression can be directly implemented in a field theoretical calculation: having fixed a renormalization scheme for the partition function, the partition function on the replica space (which has the same UV divergence structure) will inherit a renormalization scheme and thus $S_{\rm ren}$ will be determined. This assumes that the replica trick is applicable but in practice most explicit calculations of entanglement entropy in field theory do in any case make use of the replica trick. Computations of the renormalized entanglement entropy in free field theory examples will be presented elsewhere. 

There has been considerable interest recently in supersymmetric renormalization schemes for field theories on curved spaces and, in particular, in analysing how much supersymmetry is required for the partition function to be uniquely defined \cite{Gomis2015}. It would be interesting to understand the role of supersymmetry in our analysis. 

In section \ref{four} we showed that the renormalized entanglement entropy of a disk decreases under deformations of a conformal field theory by operators of dimension $3/2 < \Delta < 5/2$. Under the CHM map, this corresponds to an increase in the F quantity when one makes corresponding deformations of the theory on a three sphere; note however that these deformations do not preserve the symmetry of the $S^3$. In the companion paper \cite{Taylor1} we find analogous results for flows which are homogeneous on the three sphere. It would be interesting to understand whether these examples indeed disprove the strong version of the F theorem, or whether the flows under consideration are unphysical.

\section*{Acknowledgments}

We would like to thank Kostas Skenderis and Balt van Rees for useful comments and discussions. 
This work was supported by the Science and Technology Facilities Council (Consolidated Grant ``Exploring the Limits of the Standard Model and Beyond'').
We thank the Simons Center for partial support during the completion of this work. This project has received funding from the European Union's Horizon 2020
research and innovation programme under the Marie Sklodowska-Curie grant
agreement No 690575.

\bibliographystyle{JHEP-2}

\providecommand{\href}[2]{#2}\begingroup\raggedright\endgroup

\end{document}